%% file: main.tex
\documentclass[conference]{IEEEtran}
\IEEEoverridecommandlockouts
\usepackage[left=0.625in,right=0.625in,top=0.72in,bottom=0.99in,columnsep=0.205in]{geometry}

\usepackage{soul}

\input{preamble}

\usepackage[font=small]{caption}
\usepackage{subcaption}

\usepackage{fontawesome}
\usetikzlibrary {patterns,patterns.meta}
\usetikzlibrary{arrows.meta,positioning,calc}
\usetikzlibrary{intersections}
\usepackage{placeins}

\defineauthors[false]{gilles, jarne, ahmet, egl}

\newcommand{\reader}{\textit{reader}\xspace}
\newcommand{\Reader}{\textit{Reader}\xspace}
\newcommand{\BD}{\textit{\gls{bd}}\xspace}
\newcommand{\AZFname}{\gls{dls}} % rename here if needed
\newcommand{\AZF}{\AZFname\xspace}
\newcommand{\AZFmath}{\mathrm{\AZFname}}

\definecolor{rcolor}{RGB}{0,114,178}

\definecolor{bdcolor}{RGB}{213,94,0}

\def\PRoffset{10.67}
\def\BDoffset{8.45}

\begin{document}\sloppy

\title{Experimental Study of Interference Suppression for Backscatter Communication in Distributed MIMO
\thanks{The work of Ahmet Kaplan and Erik G. Larsson was supported in part by the Swedish Research Council (VR), in part by the Knut and Alice Wallenberg (KAW) Foundation, and in part by Excellence Center at Linköping - Lund in Information Technology (ELLIIT).}%
\thanks{This work was partially supported by the AMBIENT-6G project, which received funding from the Smart Networks and Services Joint Undertaking (SNS JU) under the European Union's Horizon Europe research and innovation programme under Grant Agreement No. 101192113.}%
\thanks{This paper is accepted at IEEE Wireless Power Technologies Conference and Expo 2026 (WPTCE 2026) and can be found in~\cite{Call2607:Experimental}.}
}

%\author{\IEEEauthorblockN{Anonymous for Review}}

\author{
\IEEEauthorblockN{Ahmet Kaplan\IEEEauthorrefmark{1}, Gilles Callebaut\IEEEauthorrefmark{2}, Jarne Van Mulders\IEEEauthorrefmark{2}, Erik G. Larsson\IEEEauthorrefmark{1}}

\IEEEauthorblockA{\IEEEauthorrefmark{1}\textit{Department of Electrical Engineering,
Linköping University}, Sweden}

\IEEEauthorblockA{\IEEEauthorrefmark{2}\textit{Department of Electrical Engineering, KU Leuven}, Belgium}
}

\maketitle

\begin{abstract}
Bistatic backscatter communication requires strong illumination of a \BD, while a spatially separated \reader detects the weak modulated reflection. In practice, the resulting \gls{dli} at the \reader can dominate the received backscattered signal and limit detection performance. This paper experimentally investigates transmit beamforming that jointly maximizes \BD illumination and suppresses \gls{dli} at the \reader in a distributed \acrlong{mimo} setup. We compare \gls{pomrt} with the proposed \AZF scheme, which enforces a spatial null at the \reader under per-antenna power constraints. Measurements using a phase-coherent \num{42}-element ceiling array at \SI{920}{\mega\hertz} show that \AZF reduces the \gls{dli} at the target \reader 
and improves the signal-to-interference ratio by up to \SI{31}{\decibel} compared to \gls{pomrt}.

\end{abstract}


% =========================
\section{Introduction}
% =========================
\glsresetall

\Gls{bc} enables ultra-low-power connectivity by encoding data in the reflection coefficient of a passive or semi-passive \gls{bd}, avoiding an active RF transmitter at the device. Early ambient and carrier-based designs demonstrated the feasibility of communicating with RF-powered devices using commodity infrastructure~\cite{Liu2013:Ambient,Kellogg2014:Wi-fi}.

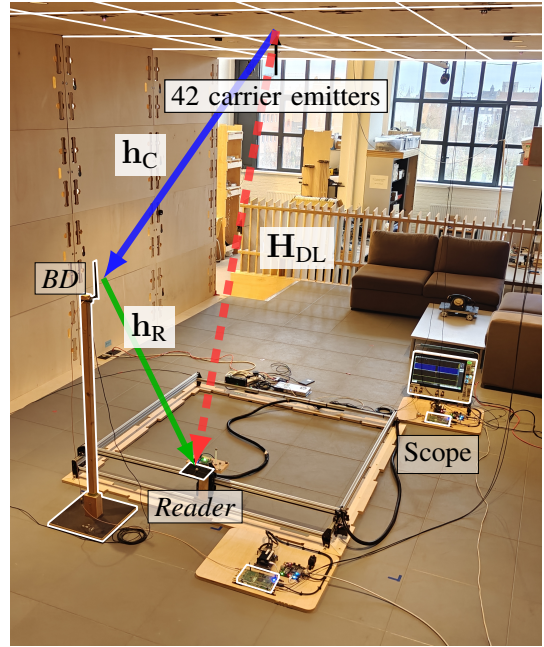
\begin{figure}[tbp]
\centering
\resizebox{0.8\linewidth}{!}{\input{figures/setup.tikz}}
\caption{Experimental setup with channels.}\label{fig:setup}
\end{figure}

\gls{bc} systems can be broadly classified into three architectures: \gls{mobc}, \gls{bibc}, and \gls{ambc}. In all cases, a \gls{bc} link consists of three key components: a \gls{ce}, a reader, and a \gls{bd}. The \gls{ce} illuminates the \gls{bd} with a continuous-wave carrier, the \gls{bd} modulates its information by switching its reflection coefficient, and the reader receives and decodes the backscattered signal.

In \gls{mobc}, the \gls{ce} and reader are collocated, resulting in a compact but less flexible architecture. In contrast, \gls{ambc} relies on ambient RF sources such as TV towers or cellular base stations, eliminating the need for a dedicated \gls{ce} but offering limited control over the illumination and link budget. Finally, \gls{bibc} employs a dedicated \gls{ce} that is spatially separated from the reader, enabling flexible deployment, improved coverage, and enhanced link budgets through independent placement of illumination and reception nodes. Another advantage is that there is no need for full deplux hardware, in contrast to \gls{mobc}.
Owing to these advantages, this work focuses on \gls{bibc}.

Three fundamental challenges in \gls{bibc} systems are i) the severe path loss of the backscatter cascade channel, ii) the strong \gls{dli} from the \gls{ce} to the \reader and iii) the backscatter efficiency. Because the backscattered signal undergoes double propagation, its power at the \reader is orders of magnitude lower than that of the direct carrier, i.e., the \gls{dli}. This power imbalance can drive the \reader front-end into non-linear operation or significantly reduce the effective number of bits of the \glspl{adc}.

Although analog \gls{dli} cancellation can mitigate this issue, it typically incurs additional hardware complexity and calibration overhead. A complementary and hardware-efficient alternative is to suppress the \gls{dli} already at the transmitter by means of multi-antenna \gls{bf} \cite{kaplan2023direct}. 
By shaping the illumination, the transmitted field can be maximized at the \gls{bd} while enforcing a spatial null at the \reader location. This spatial-domain suppression directly alleviates front-end saturation and improves the effective backscatter signal-to-interference ratio.

\subsection{Related Work}
The related work on the \gls{dli} cancellation in \gls{bibc} is reviewed here. In \cite{varshney2017lorea, lopez2023designing} and \cite{li2019capacity}, the \gls{bd} shifts the frequency of the carrier signal to separate
the backscattered signal and the \gls{dli} in the frequency domain. However, this approach increases the implementation complexity at the \gls{bd} and reduces spectral efficiency. 
In \cite{tao2021novel} and \cite{li2024code},   orthogonal codes are used to separate the \gls{dli} and the backscattered signal. However, this method requires digital signal processing after \glspl{adc} in the reader circuitry. 
In \cite{kan2023differential}, another \gls{dli} subtraction method operating in the reader in the digital domain is proposed.
These methods do not address the saturation and quantization error problem in \glspl{adc}. 
In \cite{chen2024multi}, the \gls{dli} is canceled in the analog domain, increasing the complexity. In \cite{biswas2021direct}, the impact of the \gls{dli} on the achievable communication range is analyzed. The work in \cite{li2019adaptive} analyzes the effect of the imperfect successive interference cancellation. 
In~\cite{kaplan2023direct,kaplan2025joint} and \cite{deutschmann2025physically}, transmit \gls{bf} techniques are proposed to cancel the \gls{dli} and/or focus power on the \gls{bd}. 

\subsection{Contributions}
This work provides an experimental evaluation of the transmit \gls{bf} technique proposed in \cite{kaplan2025joint},
for focusing power to the \gls{bd} while suppressing the \gls{dli} in a \gls{bibc} system with a distributed \gls{mimo} setup. 
We use a \num{42}-element phase-coherent ceiling array and implement two transmit strategies: baseline \gls{pomrt} focusing on \BD illumination and a \AZF design that constrains the \gls{dli} at the \reader. The main experimental results include:
\begin{itemize}
    \item Spatial interference heatmaps showing the superior performance of the \AZF technique to suppress the \gls{dli} at the \reader location while focusing power to the \gls{bd}.
    \item Measured BD and \reader power versus the number of active transmit elements, illustrating the trade-off between BD illumination and \gls{dli} control in practice.
\end{itemize}
All source code and data is publicly available in~\faicon{github}\footnote{\url{https://github.com/techtile-by-dramco/DLI-Supression-Experiment}}.

\input{sections/system_model}
\input{sections/algorithm}

\begin{figure*}[tbp]
    \centering

    \begin{subfigure}[t]{0.3\linewidth}
        \hfill
        \resizebox{!}{4.5cm}{\input{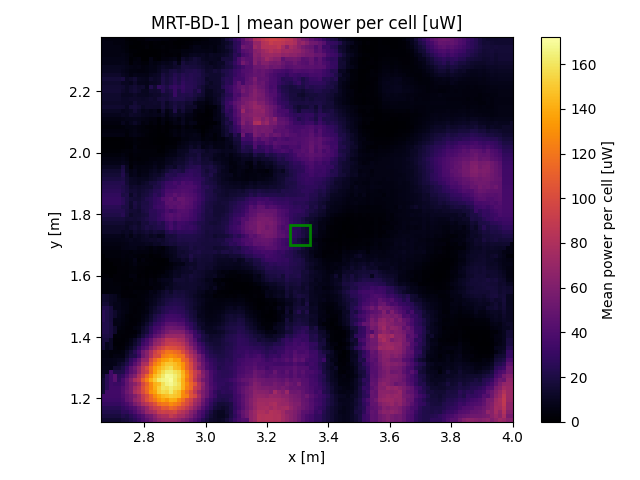}}
        \caption{PO-MRT towards the \BD showing a localized power at potential \reader positions.}
        \label{fig:heatmap-mrt-bd}
    \end{subfigure}\hfill%
    \begin{subfigure}[t]{0.3\linewidth}
        \resizebox{!}{4.5cm}{\input{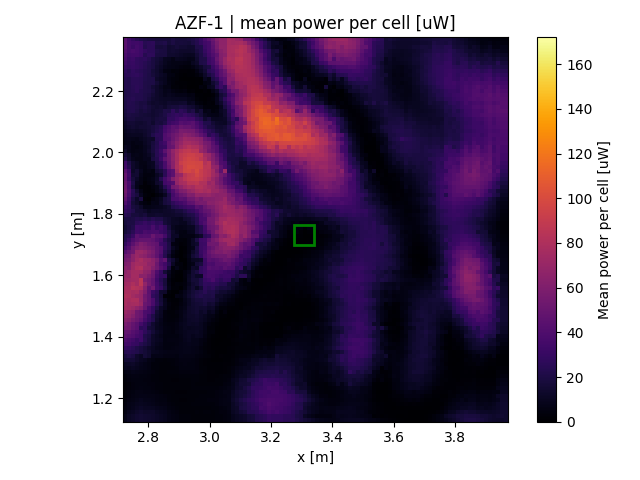}}\hfill
        \caption{\AZF is resulting in suppressed interference at the \reader location.}
        \label{fig:heatmap-azf}
    \end{subfigure}\hfill
    \begin{subfigure}[t]{0.37\linewidth}
        \centering
        \resizebox{!}{4.5cm}{\input{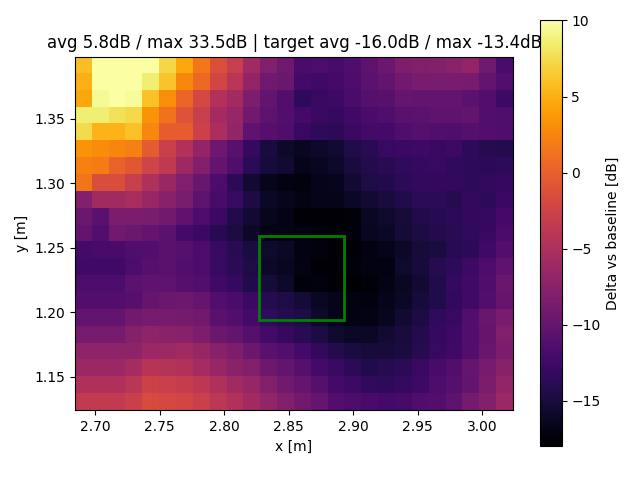}}
        \caption{Zoomed differential heatmap showing \AZF power relative to PO-MRT in \si{\decibel} (\(\lambda/2 \times \lambda/2\)).}        
        \label{fig:heatmap-azf-vs-mrt}
    \end{subfigure}

    \caption{Measured spatial distribution of received RF power over a \SI{1.25}{\meter} \(\times\) \SI{1.25}{\meter} area in \si{\micro\watt}. The dashed circle marks the \reader position and the dashed rectangle the zoom. The differential view highlights the local DLI suppression performance of \AZF relative to PO-MRT at the target location in \si{dB}.}
    \label{fig:spatial-heatmaps-beamforming}
\end{figure*}

\input{sections/exp-setup}

\begin{figure*}[hbtp]
\centering

\begin{subfigure}[t]{0.348\linewidth}
  \centering
  \input{figures/power-meas-high.tikz}
  \caption{$P_{\mathrm{BD}}$ and $P_{\mathrm{R}}$ for
  high$\rightarrow$low $\lvert h_{\mathrm{BD},m}\rvert^2$}
  \label{fig:power-meas-high}
\end{subfigure}\hfill
\begin{subfigure}[t]{0.348\linewidth}
  \centering
  \input{figures/power-meas-low.tikz}
  \caption{$P_{\mathrm{BD}}$ and $P_{\mathrm{R}}$ for low$\rightarrow$high $\lvert h_{\mathrm{BD},m}\rvert^2$}
  \label{fig:power-meas-low}
\end{subfigure}\hfill
\begin{subfigure}[t]{0.3\linewidth}
  \centering
  \input{figures/power-meas-delta.tikz}
  \caption{Power ratio $\Delta=P_{\mathrm{BD}}/P_{\mathrm{R}}$.}
  \label{fig:power-meas-delta}
\end{subfigure}

\caption{Measured power metrics versus the number of active \glspl{ce} for different CE selection methods.}\vspace{-10pt}
\label{fig:power-meas}
\end{figure*}

\input{sections/exp-results}

% =========================
\FloatBarrier%
\section{Conclusion and Future Work}
% =========================
This paper experimentally validated  large-array transmit \acrlong{bf} for  \acrlong{dli} suppression in a \acrlong{bibc} setup. Compared to a phase-only variant of \acrlong{mrt}, \AZF creates a deep spatial null at the \reader location and substantially improves the measured power ratio between \BD illumination and \reader interference. 
In our experiments,  \gls{csi} and phase coherence across the large distributed transmitter array were  sufficient to achieve significant interference suppression. 
Future work includes extending the evaluation from received-power metrics to end-to-end decoding performance, e.g., \acrlong{ber}, studying robustness against \gls{csi} errors and mobility, and investigating designs that exploit both amplitude and phase control to further improve the \acrlong{sir} under per-antenna constraints and nulling capabilities in multi-\reader scenarios.

\printbibliography%

\end{document}

%% file: preamble.tex
\usepackage{amsmath,amssymb,amsfonts}   % advanced math symbols & environments
\usepackage{textcomp}                   % extra text symbols (e.g. \textregistered)

\usepackage{graphicx}                   % include graphics (\includegraphics)
\usepackage[dvipsnames,svgnames,table]{xcolor}                     % driver-independent color extensions

\usepackage{array}                      % extended array/tabular features
\newcolumntype{H}{>{\setbox0=\hbox\bgroup}c<{\egroup}@{}} % ‘H’ column type hides content
\usepackage{booktabs}                   % \toprule, \midrule, \bottomrule
\usepackage{siunitx}                    % consistent typesetting of units & numbers
  \DeclareSIUnit{\dBm}{dBm}
  \DeclareSIUnit{\dBi}{dBi}
  \DeclareSIUnit{\dBsm}{dBsm}

\usepackage{soul}                       % underlining, striking out, highlighting (\hl)

\usepackage[backend=biber,
            minnames=1,
            doi=true,
            url=false,
            language=english,
            autolang=other,
            style=ieee]{biblatex}       % IEEE‐style bibliography
\usepackage[dvipsnames,svgnames,table]{xcolor} % custom table row colors
\usepackage[hidelinks]{hyperref}               % clickable links without colored boxes
\usepackage[capitalize]{cleveref}                          % intelligent \cref, \Cref autolabels

\usepackage{algorithmic}                % pseudocode environments

\usepackage{pgfplots}                   % high‐quality plots
  \pgfplotsset{compat=newest}
\usepgfplotslibrary{groupplots,dateplot} % extra plotting features
\usepackage{pgfplotstable}              % table support for PGFPlots
  \pgfplotsset{compat=1.18}              % specify version for stability

\usetikzlibrary{%
  patterns,          % fill patterns
  shapes.arrows,     % arrow tip libraries
  spy,               % zoom‐in—spyglass
  external,          % externalize figures
  calc,              % coordinate calculations
  shadings,          % custom shading
  shadows.blur,      % blurred shadows
  decorations.pathreplacing % decorative braces, etc.
}

\usepackage[xindy,nonumberlist]{glossaries} % acronym & glossary management
  \makenoidxglossaries            % no index file for glossaries
  \loadglsentries{abbr}           % load definitions from abbr.tex
  \glsdisablehyper                % avoid hyperlinks when no glossary is printed

\input{auto_names}                  % user-defined commands, names, etc.

\usepackage{orcidlink}
\AtBeginBibliography{\footnotesize}

\def\BibTeX{{\rm B\kern-.05em{\sc i\kern-.025em b}\kern-.08em
    T\kern-.1667em\lower.7ex\hbox{E}\kern-.125emX}}

\input{definitions/def-colors}
\input{definitions/def-tikz}

\colorlet{scenarioColor1}{color8}
\colorlet{scenarioColor2}{color8}

\colorlet{strategyColorGEO}{color1}
\colorlet{strategyColorCSI}{color2}
\colorlet{strategyColorRPS}{color3}

\newcommand{\vectr}[1]{\boldsymbol{\mathrm{#1}}}
\newcommand{\matr}[1]{\boldsymbol{\mathrm{#1}}}

\newcommand\norm[1]{\left\lVert#1\right\rVert}
\newcommand\abs[1]{\left\lvert#1\right\rvert}

\newcommand{\bhdl}{\bh_\text{DL}}
\newcommand{\bHdl}{\matr{H}_\text{DL}}
\newcommand{\bHbl}{\matr{H}_\text{BL}}

\newcommand{\bhc}{\matr{h}_\text{C}}

\newcommand{\bhr}{\matr{h}_\text{R}}

\newcommand{\bh}{\vectr{h}}

\newcommand{\bn}{\vectr{n}}

\newcommand{\bx}{\vectr{x}}

\newcommand{\by}{\vectr{y}}

\newcommand{\trp}{\mathsf{T}}

\newcommand{\complexset}[2]{ \mathbb{C}^{#1 \times #2}  }

\newcommand{\re}[1]{\operatorname{Re}\left\{#1\right\} }

\newcommand{\im}[1]{ \operatorname{Im}\left\{#1\right\} }

%% file: abbr.tex
% generated by Gilles Callebaut with the Github action: https://github.com/GillesC/action-abbreviations-latex

%%%%%%%%%%%% 2 %%%%%%%%%%%%
\newacronym{2d}{2D}{two-dimensional}
%%%%%%%%%%%% 3 %%%%%%%%%%%%
\newacronym{3d}{3D}{three-dimensional}
\newacronym{3gpp}{3GPP}{3rd generation partnership project}
\newacronym{3msk}{3MSK}{3-level minimum-shift-keying}
%%%%%%%%%%%% 4 %%%%%%%%%%%%
\newacronym{4g}{4G}{fourth-generation}
%%%%%%%%%%%% 5 %%%%%%%%%%%%
\newacronym{5g}{5G}{fifth-generation}
%%%%%%%%%%%% 6 %%%%%%%%%%%%
\newacronym{6dof}{6DoF}{six degrees of freedom}
\newacronym{6g}{6G}{sixth-generation}
%%%%%%%%%%%% A %%%%%%%%%%%%
\newacronym{abp}{ABP}{authentication by personalisation}
\newacronym{ac}{AC}{alternating current}
\newacronym{aci}{ACI}{adjacent channel interference}
\newacronym{aclr}{ACLR}{adjacent channel leakage ratio}
\newacronym{acpr}{ACPR}{adjacent channel power ratio}
\newacronym{adc}{ADC}{analog-to-digital converter}
\newacronym{adr}{ADR}{adaptive data rate}
\newacronym{aec}{AEC}{aluminum electrolytic capacitor}
\newacronym{aes}{AES}{Advanced Encryption Standard}
\newacronym{afe}{AFE}{analog front end}
\newacronym{ag}{AG}{array gain}
\newacronym{agc}{AGC}{automatic gain controller}
\newacronym{agps}{A-GPS}{Assisted Global Positioning System} %bestaat al :) % Aha, dan maak ik er wel Assisted gps van!
\newacronym{ai}{AI}{artificial intelligence}
\newacronym{alk}{ALK}{Alkaline}
\newacronym{am}{AM}{amplitude modulation}
\newacronym{amam}{AM/AM}{amplitude modulation to amplitude modulation}
\newacronym{ambc}{AmBC}{ambient BC}
\newacronym{amc}{AMC}{automatic modulation classification}
\newacronym{amp}{AMP}{approximate message passing}
\newacronym{ampm}{AM/PM}{amplitude modulation to phase modulation}
\newacronym{aoa}{AOA}{angle-of-arrival}
\newacronym{aod}{AOD}{angle-of-departure}
\newacronym{ap}{AP}{access point}
\newacronym{api}{API}{application program interface}
\newacronym{apt}{APT}{acoustic power transfer}
\newacronym{apu}{APU}{access point unit}
\newacronym{ar}{AR}{augmented reality}
\newacronym{arima}{ARIMA}{Auto-Regressive Integrated Moving Average}
\newacronym{arp}{ARP}{Antenna Reference Point}
\newacronym{asic}{ASIC}{application specific integrated circuit}
\newacronym{ask}{ASK}{amplitude-shift keying}
\newacronym{at}{AT}{ATtention}
\newacronym{auv}{AUV}{autonomous underwater vehicle}
\newacronym{awg}{AWG}{arbitrary waveform generator}
\newacronym{awgn}{AWGN}{additive white Gaussian noise}
%%%%%%%%%%%% B %%%%%%%%%%%%
\newacronym{baw}{BAW}{bulk acoustic wave}
\newacronym{bb}{BB}{base-band}
\newacronym{bc}{BC}{backscatter communication}
\newacronym{bcjr}{BCJR}{Bahl-Cocke-Jelinek-Raviv}
\newacronym{bd}{BD}{backscatter device}
\newacronym{be}{BE}{Belgium}
\newacronym{ber}{BER}{bit error rate}
\newacronym{bf}{BF}{beamforming}
\newacronym{bga}{BGA}{ball grid array}
\newacronym{bibc}{BiBC}{bistatic BC}
\newacronym{bldc}{BLDC}{brushless DC}
\newacronym{bler}{BLER}{block error rate}
\newacronym{bod}{BOD}{Brown-Out Detection}
\newacronym{bom}{BOM}{bill of materials}
\newacronym{bpsk}{BPSK}{binary phase-shift keying}
\newacronym{brp}{BRP}{beam refinement process}
\newacronym{bs}{BS}{base station}
\newacronym{bu}{BU}{booster unit}
\newacronym{bw}{BW}{bandwidth}
%%%%%%%%%%%% C %%%%%%%%%%%%
\newacronym{ca}{CA}{carrier emitter}
\newacronym{cad}{CAD}{channel activity detection}
\newacronym{cars}{CARS}{calibration reference signal}
\newacronym{cbm}{CBM}{condition based maintenance}
\newacronym{cc}{CC}{constant current}
\newacronym{ccdf}{CCDF}{complementary cumulative distribution function}
\newacronym{ccnn}{CCNN}{circular convolutional neural network}
\newacronym{ccs}{CCS}{correlative channel sounder}
\newacronym{cdf}{CDF}{cumulative distribution function}
\newacronym{cdma}{CDMA}{code division-multiple access}
\newacronym{cdrx}{CDRX}{connected mode DRX}
\newacronym{ce}{CE}{carrier emitter}
\newacronym{ced}{CED}{cumulative energy density}
\newacronym{ceofdm}{CE-OFDM}{constant-envelope OFDM}
\newacronym{cf}{CF}{cell-free}
\newacronym{cfmmimo}{CF-mMIMO}{cell-free massive MIMO}
\newacronym{cfo}{CFO}{carrier frequency offset}
\newacronym{cir}{CIR}{channel impulse response}
\newacronym{cla}{CLA}{closed-loop approach}
\newacronym{clk}{CLK}{clock}
\newacronym{cmos}{CMOS}{complementary metal oxide semiconductor}
\newacronym{cnn}{CNN}{convolutional neural network}
\newacronym{co}{CO}{combinatorial optimization}
\newacronym{cordic}{CORDIC}{coordinate rotation digital computer}
\newacronym{cost}{COST}{commercial off-the-shelf}
\newacronym{cots}{COTS}{commercial off-the-shelf}
\newacronym{cp}{CP}{cyclic prefix}
\newacronym{cpe}{CPE}{common phase error}
\newacronym{cpfsk}{CPFSK}{continuous phase frequency shift keying}
\newacronym{cpm}{CPM}{continuous phase modulation}
\newacronym{cpt}{CPT}{capacitive power transfer}
\newacronym{cpu}{CPU}{central-processing unit}
\newacronym{cpw}{CPW}{coplanar waveguide}
\newacronym{cqi}{CQI}{channel quality indicator}
\newacronym{cr}{CR}{coding rate}
\newacronym{crc}{CRC}{cyclic redundancy check}
\newacronym{crlb}{CRLB}{Cram\'er-Rao lower bound}
\newacronym{crs}{CRS}{cell reference signal}
\newacronym{cs}{CS}{compressed sensing}
\newacronym{cse}{CSE}{channel state estimation}
\newacronym{csi}{CSI}{channel state information}
\newacronym{csp}{CSP}{contact service point}
\newacronym{css}{CSS}{chirp spread spectrum}
\newacronym{cu}{CU}{central unit}
\newacronym{cv}{CV}{constant voltage}
\newacronym{cw}{CW}{continuous wave}
%%%%%%%%%%%% D %%%%%%%%%%%%
\newacronym{d2d}{D2D}{device-to-device}
\newacronym{dab}{DAB}{Digital Audio Broadcasting}
\newacronym{dac}{DAC}{digital-to-analog converter}
\newacronym{daq}{DAQ}{data acquisition system}
\newacronym{das}{DAS}{distributed antenna systems}
\newacronym{dbpsk}{DBPSK}{Differential Binary Phase Shift Keying}
\newacronym{dc}{DC}{direct current}
\newacronym{dcc}{DCC}{dynamic cooperation clustering}
\newacronym{ddc}{DDC}{digital down conversion}
\newacronym{de}{DE}{drain efficiency}
\newacronym{dft}{DFT}{discrete Fourier transform}
\newacronym{dftsofdm}{DFT-s-OFDM}{discrete Fourier Transform spread OFDM}
\newacronym{dftsofdmfdss}{DFT-s-OFDM-FDSS}{DFT-s-OFDM with frequency-domain spectral shaping}
\newacronym{dftsofdmfdssse}{DFT-s-OFDM-FDSS-SE}{DFT-s-OFDM with FDSS and spectral extension}
\newacronym{dl}{DL}{downlink}
\newacronym{dlc}{DLC}{Distributed Laser Charging}
\newacronym{dli}{DLI}{direct link interference}
\newacronym{dlt}{DLT}{Distributed Ledger Technology}
\newacronym{dm}{DM}{diffuse multipath}
\newacronym{dma}{DMA}{Direct Memory Access}
\newacronym{dmac}{DMAC}{Direct Memory Access Controller}
\newacronym{dmc}{DMC}{diffuse multipath component}
\newacronym{dmimo}{D-MIMO}{distributed MIMO}
\newacronym{dnn}{DNN}{deep neural network}
\newacronym{doa}{DOA}{direction-of-arrival}
\newacronym{dp}{DP}{differencial privacy}
\newacronym{dpd}{DPD}{digital pre-distortion}
\newacronym{dpdk}{DPDK}{Data Plane Development Kit}
\newacronym{dram}{DRAM}{dynamic random-access memory}
\newacronym{drcs}{$\Delta$RCS}{differential-radar cross section}
\newacronym{drx}{DRX}{Discontinuous Reception Mode}
\newacronym{dsb}{DSB}{double-sideband}
\newacronym{dsl}{DSL}{digital subscriber line}
\newacronym{dsp}{DSP}{digital signal processing}
\newacronym{dss}{DSS}{Dataset Storage Standard}
\newacronym{duc}{DUC}{digital up-converter}
\newacronym{dvb}{DVB}{Digital Video Broadcasting}
%%%%%%%%%%%% E %%%%%%%%%%%%
\newacronym{e2e}{E2E}{end-to-end}
\newacronym{easa}{EASA}{European Union Aviation Safety Agency}
\newacronym{ebg}{EBG}{electromagnetic bandgap}
\newacronym{ebw}{EBW}{excess bandwidth}
\newacronym{ec}{EC}{European Commission}
\newacronym{ecc}{ECC}{elliptic curve cryptography}
\newacronym{ecdf}{eCDF}{empirical cumulative distribution function}
\newacronym{ecdlp}{ECDLP}{elliptic curve discrete logarithm problem}
\newacronym{ecsp}{ECSP}{edge computing service point}
\newacronym{edlc}{EDLC}{electrostatic double-layer capacitors}
\newacronym{edrx}{eDRX}{Extended Discontinuous Reception Mode}
\newacronym{ee}{EE}{energy efficiency}
\newacronym{egprs}{EGPRS}{Enhanced Data Rates for GSM Evolution}
\newacronym{eh}{EH}{energy harvesting}
\newacronym{eirp}{EIRP}{equivalent isotropically radiated power}
\newacronym{em}{EM}{electromagnetic}
\newacronym{embb}{eMBB}{enhanced Mobile Broadband}
\newacronym{en}{EN}{energy neutral}
\newacronym{end}{END}{energy-neutral device}
\newacronym{enob}{ENOB}{effective number of bits}
\newacronym{eol}{EoL}{end of life}
\newacronym{ep}{EP}{energy profiler}
\newacronym{epd}{EPD}{electronic paper display}
\newacronym{epu}{EPU}{edge processing unit}
\newacronym{er}{ER}{Energy Receiver}
\newacronym{erp}{ERP}{effective radiation power}
\newacronym[plural=ESCs,firstplural=electronic speed controllers (ESCs)]{esc}{ESC}{electronic speed control}
\newacronym{esd}{ESD}{electrostatic discharge}
\newacronym{esl}{ESL}{electronic shelf label}
\newacronym{esr}{ESR}{equivalent series resistance}
\newacronym{et}{ET}{Energy Transmitter}
\newacronym{etsi}{ETSI}{European Telecommunications Standards Institute}
\newacronym{evd}{EVD}{eigenvalue decomposition}
\newacronym{evm}{EVM}{Error Vector Magnitude}
\newacronym{ewlb}{eWLB}{Embedded Wafer Level Ball Grid Array}
%%%%%%%%%%%% F %%%%%%%%%%%%
\newacronym{fa}{FA}{federation anchor}
\newacronym{fair}{FAIR}{Findability, Accessibility, Interoperability, and Reuse of digital assets}
\newacronym{fcf}{FCF}{frequency correlation function}
\newacronym{fd}{FD}{front-haul distance}
\newacronym{fdd}{FDD}{frequency-division duplexing}
\newacronym{fde}{FDE}{frequency domain equalizer}
\newacronym{fdm}{FDM}{frequency-division multiplexing}
\newacronym{fdma}{FDMA}{frequency division multiple access}
\newacronym{fdss}{FDSS}{frequency-domain spectral shaping}
\newacronym{fem}{FEM}{finite element analysis}
\newacronym{fembb}{feMBB}{further enhanced mobile broadband}
\newacronym{fft}{FFT}{fast Fourier transform}
\newacronym{fh}{FH}{fronthaul}
\newacronym{fhss}{FHSS}{frequency hopping spread spectrum}
\newacronym{fifo}{FIFO}{First In, First Out}
\newacronym{fim}{FIM}{Fisher information matrix}
\newacronym{fits}{FITS}{Flexible Image Transport System}
\newacronym{fl}{FL}{federated learning}
\newacronym{fom}{FoM}{figure of merit}
\newacronym{fov}{FOV}{field of view}
\newacronym{fpga}{FPGA}{field-programmable gate array}
\newacronym{fr1}{FR1}{frequency range 1}
\newacronym{fr2}{FR2}{frequency range 2}
\newacronym{fram}{FRAM}{Ferroelectric Random Access Memory}
\newacronym{fsk}{FSK}{frequency shift keying}
\newacronym{fspl}{FSPL}{free space path loss}
\newacronym{fss}{FSS}{frequency selective surface}
%%%%%%%%%%%% G %%%%%%%%%%%%
\newacronym{gb}{GB}{grant-based}
\newacronym{gdpr}{GDPR}{general data protection regulation}
\newacronym{gf}{GF}{grant-free}
\newacronym{gmsk}{GMSK}{Gaussian minimum-shift keying}
\newacronym{gnb}{gNB}{Next Generation Node B}
\newacronym{gni}{GNI}{gross national income}
\newacronym{gnn}{GNN}{graph neural network}
\newacronym{gnss}{GNSS}{global navigation satellite system}
\newacronym{gpclk}{GPCLK}{general purpose clock}
\newacronym{gpio}{GPIO}{General-Purpose Input/Output}
\newacronym{gpl}{GPL}{GNU General Public License}
\newacronym{gprs}{GPRS}{General Packet Radio Services}
\newacronym{gps}{GPS}{Global Positioning System}
\newacronym{gpu}{GPU}{graphical processing unit}
\newacronym{grc}{GRC}{GNU Radio Companion}
\newacronym{gscm}{GSCM}{geometry‐based stochastic model}
\newacronym{gsm}{GSM}{Global System for Mobile Communications}  %
\newacronym{gspm}{GSpM}{Generalized spatial modulation}
\newacronym{gwp}{GWP}{Global Warming Potential}
%%%%%%%%%%%% H %%%%%%%%%%%%
\newacronym{harq}{HARQ}{hybrid automatic repeat request}
\newacronym{hat}{HAT}{hardware attached on top}
\newacronym{hcs}{HCS}{human-centric services}
\newacronym{hdf5}{HDF5}{Hierarchical Data Format version 5}
\newacronym{hfss}{HFSS}{High Frequency Simulator Software}
\newacronym{hmd}{HMD}{head-mounted display}
\newacronym{hpbm}{HPBM}{half power beam width}
%%%%%%%%%%%% H %%%%%%%%%%%%
\newacronym{HyMPRo}{HyMPRo}{Hybrid Multi-Path Routing algorithm}
%%%%%%%%%%%% I %%%%%%%%%%%%
\newacronym{i.i.d.}{i.i.d.}{independent and identically distributed}
\newacronym{i2c}{I2C}{Inter-Integrated Circuit}
\newacronym{iaq}{IAQ}{Indoor Air Quality}
\newacronym{ib}{IB}{in-band}
\newacronym{ibbc}{IBBC}{inter-band beam configuration}
\newacronym{ibo}{IBO}{input back-off}
\newacronym{ic}{IC}{integrated circuit}
\newacronym{iccs}{ICCS}{Ilmsens correlative channel sounder}
\newacronym{ici}{ICI}{intercarrier interference}
\newacronym{icnirp}{ICNIRP}{International Commission on Non-Ionizing Radiation Protection}
\newacronym{id}{ID}{information decoding}
\newacronym{idft}{IDFT}{inverse discrete Fourier transform}
\newacronym{idxm}{IDXM}{index modulation}
\newacronym{if}{IF}{intermediate-frequency}
\newacronym{ifft}{IFFT}{inverse fast-Fourier-transform}
\newacronym{iid}{i.i.d.}{independently and identically distributed}
\newacronym{iis}{IIS}{integrated information system}
\newacronym{im}{IM}{intermodulation}
\newacronym{imd}{IMD}{intermodulation distortion}
\newacronym{imu}{IMU}{inertial measurement unit}
\newacronym{inh}{InH}{indoor hotspot office}
\newacronym{io}{IO}{input/output}
\newacronym{ioe}{IoE}{Internet of Everything}
\newacronym{iot}{IoT}{Internet of Things}
\newacronym{ipt}{IPT}{inductive power transfer}
\newacronym{ipy}{IPY}{Interventions per Year}
\newacronym{iq}{IQ}{in-phase and quadrature}
\newacronym{iqi}{IQI}{IQ imbalance}
\newacronym{ir}{IR}{infrared}
\newacronym{isi}{ISI}{intersymbol interference}
\newacronym{ism}{ISM}{industrial, scientific and medical}
\newacronym{isp}{ISP}{internet service provider}
\newacronym{itu}{ITU}{International Telecommunication Union}
%%%%%%%%%%%% J %%%%%%%%%%%%
\newacronym{jesd}{JESD}{Joint Electron Devices Engineering Council}
\newacronym{jfet}{JFET}{junction field effect transistor}
%%%%%%%%%%%% K %%%%%%%%%%%%
\newacronym{kpi}{KPI}{key performance indicator}
\newacronym{ktofdm}{KT-DFT-s-OFDM}{known-tail-DFT-s-OFDM}
\newacronym{kvi}{KVI}{key value indicator}
%%%%%%%%%%%% L %%%%%%%%%%%%
\newacronym{larva}{LARVA}{LARge Virtual Array}
\newacronym{lca}{LCA}{life cycle assessment}
\newacronym{lco}{LCO}{lithium cobalt oxide}
\newacronym{ldo}{LDO}{Low-dropout voltage regulator}
\newacronym{ldpc}{LDPC}{low-density parity-check}
\newacronym{led}{LED}{Light Emitting Diode}
\newacronym{less}{LESS}{Low Energy Scheduler Solution}
\newacronym{lfp}{LFP}{lithium iron phosphate}
\newacronym{lib}{LIB}{Lithium-Ion Battery}
\newacronym{lic}{LIC}{lithium-ion capacitor}
\newacronym{lid}{LID}{Lithium Iron Disulfide}
\newacronym{lidar}{LiDAR}{light detection and ranging}
\newacronym{liion}{Li-ion}{lithium-ion}
\newacronym{lipo}{LiPo}{lithium polymer}
\newacronym{lis}{LIS}{large intelligent surface}
\newacronym{llh}{LLH}{log-likelihood}
\newacronym{lls}{LLS}{link-level simulation}
\newacronym{lmd}{LMD}{Lithium Manganese Dioxide}
\newacronym{lmmse}{LMMSE}{least minimum mean square error}
\newacronym{lmo}{LMO}{lithium ion manganese oxide}
\newacronym{lna}{LNA}{low-noise amplifier}
\newacronym{lo}{LO}{local oscillator}
\newacronym{lora}{LoRa}{long range}
\newacronym{lorawan}{LoRaWAN}{long-range wide-area network}
\newacronym{los}{LoS}{line-of-sight}
\newacronym{lp}{LP}{linear programming}
\newacronym{lpf}{LPF}{low-pass filter}
\newacronym{lpt}{LPT}{laser power transfer}
\newacronym{lpwa}{LPWA}{Low Power Wide Area}
\newacronym{lpwan}{LPWAN}{low-power wide-area network}
\newacronym{lpwans}{LPWANs}{Low-Power Wide-Area Networks}
\newacronym{lqi}{LQI}{link quality indicator}
\newacronym{lrelu}{LReLU}{leaky rectified linear unit}
\newacronym{lrt}{LRT}{likelihood-ratio test}
\newacronym{ls}{LS}{least squares}
\newacronym{lsa}{LSA}{large synthetic array}
\newacronym{lsf}{LSF}{large-scale fading}
\newacronym{lsfc}{LSFC}{large-scale fading component}
\newacronym{lstm}{LSTM}{Long Short-Term Memory}
\newacronym{ltc}{LTC}{lithium thionyl chloride}
\newacronym{lte}{LTE}{Long Term Evolution}
\newacronym{lti}{LTI}{linear time-invariant}
\newacronym{lto}{LTO}{lithium titanate}
\newacronym{lusta}{LUSTA 5G }{Logistique mUltimodale Sécuritaire Téléopérée \& Autonome 5G}
%%%%%%%%%%%% M %%%%%%%%%%%%
\newacronym{m2m}{M2M}{machine to machine}
\newacronym{mac}{MAC}{Medium Access Control}
\newacronym{mate}{MATE}{millimeter-wave MIMO testbed}
\newacronym{mc}{MC}{Monte Carlo}
\newacronym{mcl}{MCL}{Maximum Coupling Loss}
\newacronym{mcs}{MCS}{modulation and coding scheme}
\newacronym{mcu}{MCU}{microcontroller unit}
\newacronym{mec}{MEC}{multi-access edge computing}
\newacronym{mems}{MEMS}{micro-electromechanical systems}
\newacronym{mf}{MF}{matched filter}
\newacronym{mimo}{MIMO}{multiple-input multiple-output}
\newacronym{miso}{MISO}{multiple-input single-output}
\newacronym{ml}{ML}{machine learning}
\newacronym{mlp}{MLP}{multilayer perceptron}
\newacronym{mmic}{MMIC}{monolithic microwave integrated circuit}
\newacronym{mmimo}{mMIMO}{massive MIMO}
\newacronym{mmse}{MMSE}{minimum mean square error}
\newacronym{mmtc}{mMTC}{massive machine-typed communication}
\newacronym{mmwave}{mmWave}{millimeter wave}
\newacronym{mn}{MN}{matching network}
\newacronym{mobc}{MoBC}{monostatic BC}
\newacronym{mosfet}{MOSFET}{metal-oxide semiconductor field effect transistor}
\newacronym{mpc}{MPC}{multipath component}
\newacronym{mppt}{MPPT}{maximum power point tracking}
\newacronym{mr}{MR}{maximum ratio}
\newacronym{mrc}{MRC}{maximum ratio combining}
\newacronym{mrc_em}{MRC}{maximum ratio combining}
\newacronym{mrc_EM}{MRC}{Magnetic Resonance Coupling}
\newacronym{mrt}{MRT}{maximum ratio transmission}
\newacronym{mse}{MSE}{mean square error}
\newacronym{msk}{MSK}{Minimum-Shift Keying}
\newacronym{mtc}{MTC}{Machine-Type Communication}
\newacronym{multi-rat}{Multi-RAT}{multiple radio access technology}
\newacronym{multirat}{Multi-RAT}{Multiple Radio Access Technology}
\newacronym{music}{MUSIC}{MUltiple SIgnal Classification}
%%%%%%%%%%%% N %%%%%%%%%%%%
\newacronym{navauwall}{NAVAUWALL}{AUtomated NAVigation in WALLonia}
\newacronym{nb}{NB}{narrowband}
\newacronym{nbiot}{NB-IoT}{narrowband IoT}
\newacronym{nca}{NCA}{nickel cobalt aluminum}
\newacronym{netcdf}{NetCDF}{Network Common Data Form}
\newacronym{nf}{NF}{noise figure}
\newacronym{nfc}{NFC}{near-field communication}
\newacronym{nfv}{NFV}{network function virtualization}
\newacronym{ngmn}{NGMN}{Next Generation Mobile Networks }
\newacronym{ni}{NI}{National Instruments}
\newacronym{nicd}{NiCd}{nikkel cadmium}
\newacronym{nimh}{NiMH}{nikkel metal hydride}
\newacronym{nlos}{NLoS}{non-line-of-sight}
\newacronym{nmc}{NMC}{nickel manganese cobalt}
\newacronym{nmos}{nMOS}{n-channel metal-oxide semiconductor}
\newacronym{nn}{NN}{neural network}
\newacronym{nnls}{NNLS}{non-negative least squares}
\newacronym{noma}{NOMA}{non-orthogonal multiple access}
\newacronym{np}{NP}{Neyman-Pearson}
\newacronym{npbch}{NPBCH}{Narrowband Physical Broadcast Channel}
\newacronym{npss}{NPSS}{Narrow Band Primay Synchronization Signal}
\newacronym{nr}{NR}{New Radio}
\newacronym{nrs}{NRS}{Narrow Band Reference Signal}
\newacronym{nsss}{NSSS}{Narrowband Secondary Synchronization Signal}
\newacronym{ntp}{NTP}{network time protocol}
%%%%%%%%%%%% O %%%%%%%%%%%%
\newacronym{oai}{OAI}{OpenAirInterface} % no spelling mistake, it is spelled all glued to eachother...
\newacronym{obw}{OBW}{occupied bandwidth}
\newacronym{ofdm}{OFDM}{orthogonal frequency-division multiplexing}
\newacronym{ofdma}{OFDMA}{orthogonal frequency-division multiple access}
\newacronym{ofdmim}{OFDM-IM}{OFDM with index modulation}
\newacronym{olos}{OLoS}{obstructed-line-of-sight}
\newacronym{oma}{OMA}{orthogonal multiple access}
\newacronym{oob}{OOB}{out-of-band}
\newacronym{ook}{OOK}{on-off keying}
\newacronym{opbo}{OPBO}{output power backoff}
\newacronym{oran}{O-RAN}{open radio-access network}
\newacronym{os}{OS}{operating system}
\newacronym{ota}{OTA}{over-the-air}
\newacronym{otaa}{OTAA}{over-the-air authentication}
%%%%%%%%%%%% P %%%%%%%%%%%%
\newacronym{p1}{P1}{Phase 1}
\newacronym{p2}{P2}{Phase 2}
\newacronym{p2p}{P2P}{point-to-point}
\newacronym{pa}{PA}{power amplifier}
\newacronym{pae}{PAE}{power-added efficiency}
\newacronym{pam}{PAM}{pulse amplitude modulation}
\newacronym{pana}{PanA}{Panel A}
\newacronym{panb}{PanB}{Panel B}
\newacronym{papr}{PAPR}{peak-to-average power ratio}
\newacronym{pc}{PC}{pilot count}
\newacronym{pcb}{PCB}{printed circuit board}
\newacronym{pcg}{PCG}{power consumption gain}
\newacronym{pcie}{PCIe}{Peripheral Component Interconnect Express}
\newacronym{pcsi}{PCSI}{perfect channel state information}
\newacronym{pd}{PD}{powered device}
\newacronym{pdcch}{PDCCH}{physical downlink control channel}
\newacronym{pdf}{PDF}{probability density function}
\newacronym{pdp}{PDP}{power delay profile}
\newacronym{pdsch}{PDSCH}{physical downlink shared channel}
\newacronym{pe}{PE}{processing element}
\newacronym{peb}{PEB}{positioning error bound}
\newacronym{per}{PER}{packet error rate}
\newacronym{pet}{PET}{privacy enhancing technology}
\newacronym{pg}{PG}{path gain}
\newacronym{pgd}{PGD}{proximal gradient descent}
\newacronym{phy}{PHY}{physical}
\newacronym{pki}{PKI}{public key infrastucture}
\newacronym{pl}{PL}{path loss}
\newacronym{pla}{PLA}{physically large array}
\newacronym{pll}{PLL}{phase-locked loop}
\newacronym[plural=PMs,firstplural=person months (PMs)]{pm}{PM}{person month}
\newacronym{pmf}{PMF}{polymer microwave fiber}
\newacronym{pmu}{PMU}{power management unit}
\newacronym{pn}{PN}{pseudo-noise}
\newacronym{po}{PO}{phase offset}
\newacronym{poc}{PoC}{proof of concept}
\newacronym{poe}{PoE}{power-over-Ethernet}
\newacronym{pps}{1PPS}{pulse per second}
\newacronym{pr}{PR}{phase reversal}
\newacronym{prb}{PRB}{Physical Resource Block}
\newacronym{prbs}{PRBs}{Physical Resource Blocks}
\newacronym{prs}{PRS}{Peripheral Reflex System}
\newacronym{ps}{PS}{Processing System}
\newacronym{psd}{PSD}{power spectral density}
\newacronym{pse}{PSE}{power sourcing equipment}
\newacronym{psk}{PSK}{phase shift keying}
\newacronym{psm}{PSM}{power saving mode}
\newacronym{pss}{PSS}{primary synchronisation signal}
\newacronym{ptp}{PTP}{precision-time protocol}
\newacronym{ptrs}{PTRS}{Phase-Tracking Reference Signals}
\newacronym{ptw}{PTW}{paging time window}
\newacronym{pv}{PV}{photovoltaic}
\newacronym{pw}{PW}{planar wavefront}
\newacronym{pwm}{PWM}{pulse width modulation}
%%%%%%%%%%%% Q %%%%%%%%%%%%
\newacronym{qam}{QAM}{quadrature amplitude modulation}
\newacronym{qos}{QoS}{quality-of-service}
\newacronym{qpsk}{QPSK}{quadrature phase-shift keying}
\newacronym{qrrls}{QR-RLS}{QR decomposition based recursive least squares}
\newacronym{quadriga}{QuaDRiGa}{QUAsi Deterministic RadIo channel GenerAtor}
%%%%%%%%%%%% R %%%%%%%%%%%%
\newacronym{ra}{RA}{Random Access}
\newacronym{ram}{RAM}{random-access memory}
\newacronym{ran}{RAN}{radio access network}
\newacronym{rar}{RAR}{Random Access Response}
\newacronym{rat}{RAT}{radio access technology}
\newacronym{rb}{Rb}{Rubidium}
\newacronym{rbs}{RBS}{radio base station}
\newacronym{rbw}{RBW}{resolution bandwidth}
\newacronym{rc}{RC}{raised-cosine}
\newacronym{rcs}{RCS}{radar cross section}
\newacronym{rdl}{RDL}{redistribution layer}
\newacronym{re}{RE}{radio element}
\newacronym{relu}{ReLU}{rectified linear unit}
\newacronym{rf}{RF}{radio frequency}
\newacronym{rfeh}{RFEH}{radio frequency energy harvesting}
\newacronym{rfic}{RFIC}{radio-frequency integrated circuit}
\newacronym{rfid}{RFID}{radio frequency identification}
\newacronym{rfpt}{RFPT}{radio frequency power transfer}
\newacronym{rfsoc}{RFSoC}{Radio Frequency System-on-Chip}
\newacronym{rir}{RIR}{room impulse response}
\newacronym{ris}{RIS}{reflective intelligent surface}%reconfigurable?
\newacronym{rllmtc}{RLLMTC}{reliable low latency machine type communication}
\newacronym{rls}{RLS}{recursive least squares}
\newacronym{rms}{RMS}{root-mean-square}
\newacronym{rmse}{RMSE}{root-mean-square error}
\newacronym{rmt}{RMT}{random matrix theory}
\newacronym{rof}{RoF}{radio-over-fiber}
\newacronym{ros}{ROS}{robot operating system}
\newacronym{rpi}{RPi}{Raspberry Pi}
\newacronym{rrc}{RRC}{Radio Resource Connection}
\newacronym{rreq}{RREQ}{route request packet}
\newacronym{rsrp}{RSRP}{Reference Signals Received Power}
\newacronym{rsrq}{RSRQ}{Reference Signal Received Quality}
\newacronym{rss}{RSS}{received signal strength}
\newacronym{rssi}{RSSI}{received signal strength indicator}
\newacronym{rtc}{RTC}{real time clock}
\newacronym{rtf}{RTF}{reader talks first}
\newacronym{rtk}{RTK}{real time kinematics}
\newacronym{rts}{RTS}{ray tracing simulator}
\newacronym{ru}{RU}{Radio Unit}
\newacronym{rv}{RV}{random variable}
\newacronym{rw}{RW}{RadioWeaves}
\newacronym{rx}{RX}{receiver}
\newacronym{rzf}{RZF}{regularized zero forcing}
%%%%%%%%%%%% S %%%%%%%%%%%%
\newacronym{s-parameter}{S-parameter}{scattering parameter}
\newacronym{sa}{SA}{synchronization anchor}
\newacronym{sa5}{SA}{Stand Alone}
\newacronym{sar}{SAR}{specific absorption rate}
\newacronym{sbl}{SBL}{sparse Bayesian learning}
\newacronym{sc}{SC}{single carrier}
\newacronym{scfde}{SC-FDE}{single-carrier modulation with frequency-domain-equalization}
\newacronym{scfdma}{SCFDMA}{single-carrier frequency division multiple access}
\newacronym{scs}{SCS}{sub-carrier spacing}
\newacronym{sdg}{SDG}{Sustainable Development Goal}
\newacronym{sdm}{SDM}{sigma-delta modulator}
\newacronym{sdma}{SDMA}{spatial-division multiple access}
\newacronym{sdn}{SDN}{software-defined network}
\newacronym{sdof}{SDoF}{sigma-delta over fiber}
\newacronym{sdr}{SDR}{software-defined radio}
\newacronym{se}{SE}{spectral efficiency}
\newacronym{sei}{SEI}{specific emitter identification}
\newacronym{ser}{SER}{symbol-error rate}
\newacronym{sf}{SF}{spreading factor}
\newacronym{sfn}{SFN}{single frequency network}
\newacronym{sfp}{SFP}{small form-factor pluggable}
\newacronym{sha}{SHA}{Secure Hash Algorithm}
%%%%%%%%%%%% S %%%%%%%%%%%%
\newacronym{SigMF}{SigMF}{Signal Metadata Format}
%%%%%%%%%%%% S %%%%%%%%%%%%
\newacronym{simo}{SIMO}{single-input multiple-output}
\newacronym{sinr}{SINR}{signal-to-interference-plus-noise ratio}
\newacronym{sir}{SIR}{signal-to-interference ratio}
\newacronym{siso}{SISO}{single-input single-output}
\newacronym{slam}{SLAM}{simultaneous localization and mapping}
\newacronym{slc}{SLC}{spatial leakage suppression}
\newacronym{slerp}{SLERP}{spherical linear interpolation}
\newacronym{sma}{SMA}{SubMiniature version A}
\newacronym{smc}{SMC}{specular multipath component}
\newacronym{smps}{SMPS}{switched mode power supply}
\newacronym{sndr}{SNDR}{signal-to-noise-and-distortion ratio}
\newacronym{snidr}{SNIDR}{signal-to-noise-and-interference-and-distortion ratio}
\newacronym{snir}{SNIR}{signal-to-interference-plus-noise ratio}
\newacronym{snr}{SNR}{signal-to-noise ratio}
\newacronym{soc}{SoC}{state of charge}
%%%%%%%%%%%% S %%%%%%%%%%%%
\newacronym{SoC}{SOC}{System on Chip}
%%%%%%%%%%%% S %%%%%%%%%%%%
\newacronym{sp}{SP}{service point}
\newacronym{spdt}{SPDT}{single pole double throw}
\newacronym{spi}{SPI}{Serial Peripheral Interface}
\newacronym{spst}{SPST}{single pole single throw}
\newacronym{sram}{SRAM}{static random-access memory}
\newacronym{srd}{SRD}{short-range device}
\newacronym{srls}{SRLS}{standard recursive least squares}
\newacronym{srs}{SRS}{Sounding Reference Signal}
\newacronym{ssb}{SSB}{synchronisation signal block}
\newacronym{ssd}{SSD}{solid state drive}
\newacronym{ssq}{SSQ}{simulator sickness questionnaire}
\newacronym{steam}{STEAM}{science, technology, engineering, the arts, and mathematics}
\newacronym{svd}{SVD}{singular value decomposition}
\newacronym{sw}{SW}{spherical wavefront}
\newacronym{swipt}{SWIPT}{simultaneous wireless information and power transfer}
\newacronym{synce}{SyncE}{Synchronous Ethernet}
%%%%%%%%%%%% T %%%%%%%%%%%%
\newacronym{ta}{TA}{timing advance}
\newacronym{tau}{TAU}{tracking area update}
\newacronym{tcer}{TCER}{transported to consumed energy ratio}
\newacronym{tcp}{TCP}{Transmission Control Protocol}
\newacronym{tcxo}{TCXO}{temperature compensated crystal oscillator}
\newacronym{tdce}{TD-CE}{time-domain compression and expansion}
\newacronym{tdd}{TDD}{time division duplexing}
\newacronym{tdma}{TDMA}{time division-multiple access}
\newacronym{tdoa}{TDOA}{time-difference-of-arrival}
\newacronym{thz}{THz}{Terahertz}
\newacronym{to}{TO}{timing offset}
\newacronym{toa}{TOA}{time-of-arrival}
\newacronym{tof}{ToF}{time-of-flight}
\newacronym{tosm}{TOSM}{through-open-short-match}
\newacronym{tpms}{TPMS}{Tire-Pressure Monitoring System}
\newacronym{trl}{TRL}{technology readyness level}
\newacronym{trp}{TRP}{Transmission Reception Point}
\newacronym{tsn}{TSN}{time-sensitive networking}
\newacronym{ttf}{TTF}{tag talks first}
\newacronym{ttff}{TTFF}{Time To First Fix}
\newacronym{tti}{TTI}{transmission time interval}
\newacronym{ttm}{TTM}{time to market}
\newacronym{ttn}{TTN}{The Things Network}
\newacronym{tx}{TX}{transmitter}
%%%%%%%%%%%% U %%%%%%%%%%%%
\newacronym{uart}{UART}{Universal Asynchronous Receiver/Transmitter}
\newacronym{uav}{UAV}{unmanned aerial vehicle}
\newacronym{uc}{UC}{use case}
\newacronym{ucie}{UCIe}{Universal Chiplet Interconnect Express}
\newacronym{udp}{UDP}{User Datagram Protocol}
\newacronym{ue}{UE}{user equipment}
\newacronym{ugv}{UGV}{unmanned ground vehicle}
\newacronym{uhd}{UHD}{USRP hardware driver}
\newacronym{uhf}{UHF}{ultra-high frequency}
\newacronym{ul}{UL}{uplink}
\newacronym{ula}{ULA}{uniform linear array}
\newacronym{uMIMO}{$\mu$-MIMO}{ultra-massive Multiple-Input Multiple-Output}
\newacronym{ummtc}{umMTC}{ultra massive machine type communication}
\newacronym{un}{UN}{United Nations}
\newacronym{unow}{uNOW}{unified non-orthogonal waveform}
\newacronym{upa}{UPA}{uniform planar array}
\newacronym{ura}{URA}{uniform rectangular array}
\newacronym{urllc}{URLLC}{ultra-reliable low-latency communications}
\newacronym{usrp}{USRP}{universal software radio peripheral}
\newacronym{uv}{UV}{unmanned vehicle}
\newacronym{uw}{UW}{unique word}
\newacronym{uwb}{UWB}{ultrawideband}
\newacronym{uwofdm}{UW-DFT-s-OFDM}{unique word DFT-s-OFDM}
%%%%%%%%%%%% V %%%%%%%%%%%%
\newacronym{v2v}{V2V}{vehicle-to-vehicle}
\newacronym{vco}{VCO}{voltage-controlled oscillator}
\newacronym{vep}{VEP}{virtual edge platform}
\newacronym{vlc}{VLC}{visible light communication}
\newacronym{vlp}{VLP}{visible light positioning}
\newacronym{vna}{VNA}{vector network analyzer}
\newacronym{voc}{VOC}{Voltatile Organic Compound}
\newacronym{votable}{VOTable}{Virtual Observatory Table}
\newacronym{vr}{VR}{virtual reality}
\newacronym{vuca}{VUCA}{volatile, uncertain, complex and ambiguous}
%%%%%%%%%%%% W %%%%%%%%%%%%
\newacronym{wb}{WB}{wideband}
\newacronym{wban}{WBAN}{wireless body area network}
\newacronym{wd}{WD}{Wireless distance}
\newacronym{wimax}{WiMAX}{Worldwide Interoperability for Microwave Access}
\newacronym{wlan}{WLAN}{wireless LAN}
\newacronym[plural=WPs,firstplural=work packages (WPs)]{wp}{WP}{work package}
\newacronym{wpt}{WPT}{wireless power transfer}
\newacronym{wr}{WR}{White Rabbit}
\newacronym{wrsn}{WRSN}{wireless rechargeable sensor network}
\newacronym{wsn}{WSN}{wireless sensor network}
%%%%%%%%%%%% X %%%%%%%%%%%%
\newacronym{xets}{XETS}{cross exponentially tapered slot}
\newacronym{xlmimo}{XL-MIMO}{extremely large-scale MIMO}
\newacronym{xr}{XR}{extended reality}
%%%%%%%%%%%% Z %%%%%%%%%%%%
\newacronym{z3ro}{Z3RO}{zero third-order distortion}
\newacronym{zf}{ZF}{zero-forcing}
\newacronym{zmcscg}{ZMCSCG}{zero mean circularly symmetric complex Gaussian}
\newacronym{zmq}{ZMQ}{ZeroMQ}
\newacronym{ztofdm}{ZT-DFT-s-OFDM}{zero-tail DFT-s-OFDM}
\newacronym{llr}{LLR}{log-likelihood ratio}
\newacronym{rps}{RPS}{random-phase sweeping}

\newacronym{dls}{DLS}{direct-link suppression}
\newacronym{pomrt}{PO-MRT}{phase-only maximum ratio transmission}

%% file: auto_names.tex
\makeatletter
\PassOptionsToPackage{dvipsnames,svgnames,table}{xcolor}%
\makeatother

\usepackage{listofitems}
\usepackage{pgffor}
\usepackage{xparse}
\usepackage{ifthen}
\usepackage{tikz}
\usepackage{xspace}
\usetikzlibrary{external}

\makeatletter
\def\myColorList{LimeGreen,BrickRed,Fuchsia,Bittersweet,YellowOrange, YellowGreen, WildStrawberry, Fuchsia, RedViolet, Periwinkle, PineGreen, OrangeRed, RawSienna, Turquoise, Aquamarine, BlueViolet, BurntOrange, DarkOrchid, TealBlue, Violet, RoyalPurple,%
LimeGreen,BrickRed,Fuchsia,Bittersweet,YellowOrange, YellowGreen, WildStrawberry, Fuchsia, RedViolet, Periwinkle, PineGreen, OrangeRed, RawSienna, Turquoise, Aquamarine, BlueViolet, BurntOrange, DarkOrchid, TealBlue, Violet, RoyalPurple}
\readlist\ColorList\myColorList

\definecolor{defaultColor}{RGB}{112,45,128}
\NewDocumentCommand{\roundLabel}{O{defaultColor} m}{%
\tikzexternaldisable
    \tikz[baseline=(char.base)]{
        \node[rectangle, rounded corners=0.65mm, inner sep=0.65mm, fill=#1, draw=#1, text=white, font=\itshape](char) {#2};
    }%
\tikzexternalenable
}

\NewDocumentCommand{\defineauthors}{ O{true} m }{%
    \ifthenelse{\equal{#1}{true}}{%
        Personal commands for comments are:
        \begin{itemize}
        \foreach \x [count=\xi from 1] in {#2} {   
            \item \textcolor{\ColorList[\xi]}{\textbackslash\unexpanded\expandafter{\x}\{\}}
            }
        \end{itemize}
        \foreach \x [count=\xi from 1] in {#2} { 
            \expandafter\xdef\csname\x\endcsname####1{\noexpand\textcolor{\ColorList[\xi]}{\roundLabel[\ColorList[\xi]]{\unexpanded\expandafter{\x}} ####1}}%
        }
    }{%
        \foreach \x [count=\xi from 1] in {#2} { 
            \expandafter\xdef\csname\x\endcsname####1{\noexpand\textcolor{\ColorList[\xi]}{\roundLabel[\ColorList[\xi]]{\unexpanded\expandafter{\x}} ####1}}%
        }
    }
    \foreach \x [count=\xi from 1] in {#2} {
        \expandafter\xdef\csname at\x\endcsname####1{\noexpand\textcolor{\ColorList[\xi]}{@\x}}%
    }
}

\NewDocumentCommand{\removeauthors}{ m }{%
    \foreach \x in {#1} { 
        \expandafter\edef\csname\x\endcsname####1{}%
    }
    \foreach \x in {#1} {
        \expandafter\edef\csname at\x\endcsname####1{}%
    }
}

\makeatother

%% file: definitions/def-colors.tex
\definecolor{color1}{HTML}{1b9e77}
\definecolor{color2}{HTML}{d95f02}
\definecolor{color3}{HTML}{7570b3}
\definecolor{color4}{HTML}{e7298a}
\definecolor{color5}{HTML}{66a61e}
\definecolor{color6}{HTML}{e6ab02}
\definecolor{color7}{HTML}{a6761d}
\definecolor{color8}{HTML}{666666}
\colorlet{AMBgreen}{color1}
\colorlet{AMBdarkgreen}{color1}
\colorlet{AMBlightgreen}{color1}
\definecolor{darkgray176}{RGB}{176,176,176}

%% file: definitions/def-tikz.tex
\usepackage{tikz}
\usepackage{pgfplots}   % needed regardless of exteralization
\pgfplotsset{compat=newest}
\usetikzlibrary{plotmarks}
\usetikzlibrary{arrows.meta}
\usetikzlibrary{arrows}
\usetikzlibrary{calc,fadings,decorations.pathreplacing}
\usetikzlibrary{fit,backgrounds}  % for the fit of boxes

\usetikzlibrary{shapes,arrows,fit,positioning}
\usetikzlibrary{shapes}
\usetikzlibrary{spy}
\usetikzlibrary{circuits}
\usetikzlibrary{arrows}

\pgfplotsset{
    every axis/.append style={
        title style={draw=none},
        label style={font=\small},
        tick label style={font=\small},
        legend style={
            fill opacity=0.8,
            nodes={scale=0.8, transform shape}, {draw=none}
        },
        tick align=outside,
        tick pos=left,
        x grid style={darkgray176},
        xtick style={color=black},
        y grid style={darkgray176},
        ytick style={color=black},
        grid=both,
        xlabel style={yshift=-6pt},
    },
    every axis plot/.append style={
        line width=1.0pt,
    },
}

\tikzset{%
  >=latex,
  inner sep=0pt,%
  outer sep=2pt,%
  mark coordinate/.style={inner sep=0pt,outer sep=0pt,minimum size=3pt,
  fill=black,circle}%
}

\pgfplotsset{colormap={inferno}{
    rgb(0)=(0.001462, 0.000466, 0.013866),
    rgb(15)=(0.037668, 0.025921, 0.132232),
    rgb(30)=(0.116656, 0.047574, 0.272321),
    rgb(45)=(0.217949, 0.036615, 0.383522),
    rgb(60)=(0.316282, 0.053490, 0.425116),
    rgb(75)=(0.410113, 0.087896, 0.433098),
    rgb(90)=(0.503493, 0.121575, 0.423356),
    rgb(105)=(0.596940, 0.154848, 0.398125),
    rgb(120)=(0.688653, 0.192239, 0.357603),
    rgb(135)=(0.775059, 0.239667, 0.303526),
    rgb(150)=(0.851384, 0.302260, 0.239636),
    rgb(165)=(0.912966, 0.381636, 0.169755),
    rgb(180)=(0.956852, 0.475356, 0.094695),
    rgb(195)=(0.981895, 0.579392, 0.026250),
    rgb(210)=(0.987464, 0.690366, 0.079990),
    rgb(225)=(0.973088, 0.805409, 0.216877),
    rgb(240)=(0.947594, 0.917399, 0.410665),
    rgb(255)=(0.988362, 0.998364, 0.644924),
    }} 

%% file: figures/setup.tikz
\begin{tikzpicture}[
box/.style={
    fill=white, fill opacity=0.45, text opacity=1,inner sep=0.1cm, draw=black
  },
  hc/.style={
    ->,
    line width=3pt,
    draw=blue!90,
    draw opacity=0.9,inner sep=0.1cm
  },
  hr/.style={
    ->,
    line width=2.5pt,
    draw=green!70!black,
    draw opacity=0.9,inner sep=0.1cm
  },
  hdl/.style={
    ->,
    line width=3.5pt,
    draw=red!80,
    dashed,
    dash pattern=on 6pt off 4pt,
    draw opacity=0.8,inner sep=0.1cm
  }
]
  % Base image
  \node[anchor=south west, inner sep=0] (img) at (0,0)
  % trim = left bottom right top
    {\includegraphics[width=0.9\linewidth,trim=150mm 100mm 100mm 30mm,
    clip]{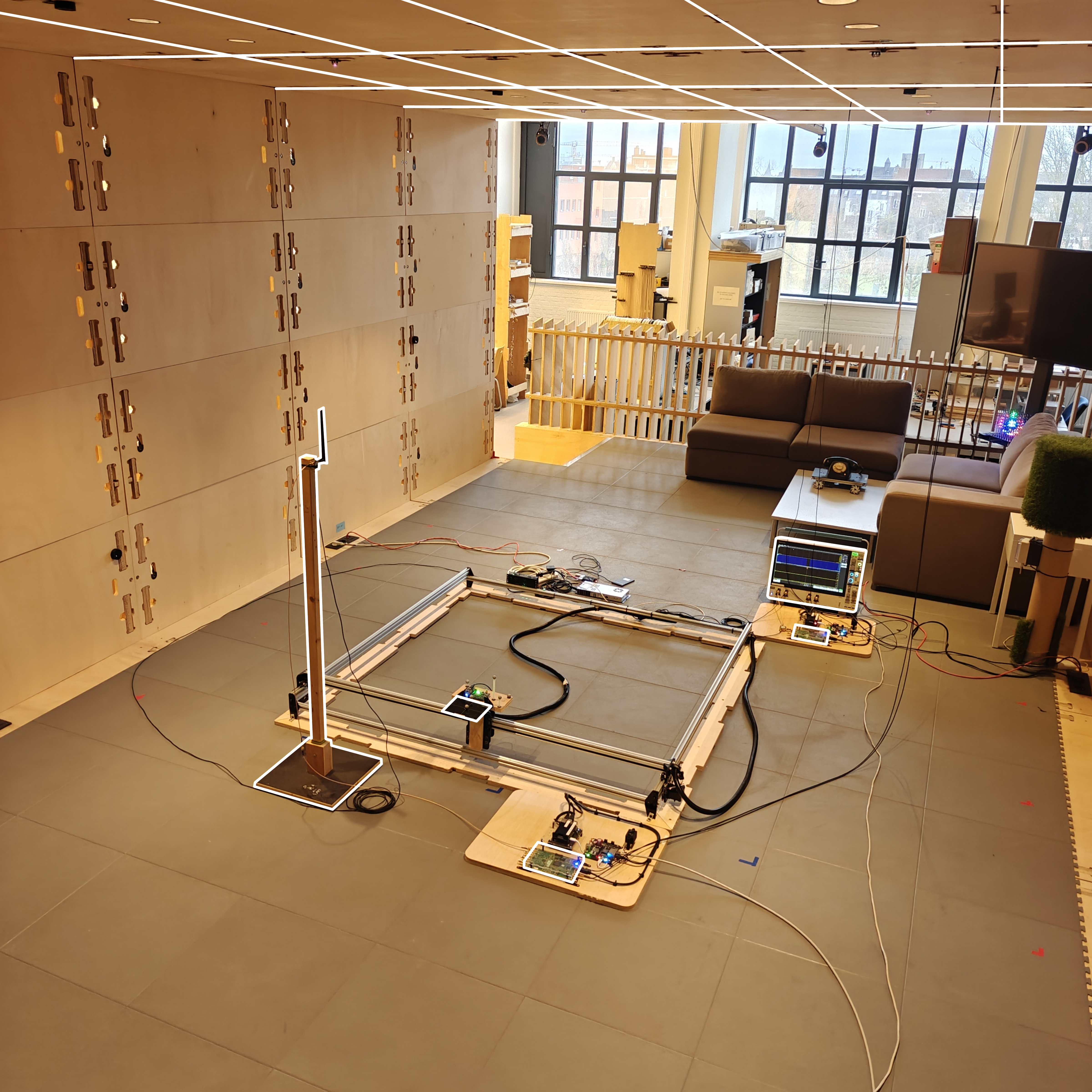}};

  % Coordinate system tied to image
  \begin{scope}[x={(img.south east)}, y={(img.north west)}]
    % Text overlays (normalized coordinates 0..1)
   
    \node[anchor=center, box, yshift=-0.5cm] (CE)
      at (0.5,0.9) {\large 42 \acrlongpl{ce}};

     \draw[->, very thick] (CE.north) -- (0.5,0.95);

    \node[anchor=center, box,] (BD) at (0.1,0.57) {\large \BD};
    \node[anchor=center, box,fill opacity=0.6] (reader) at (0.35,0.22) {\large \Reader};
    \node[anchor=center, box,] (scope) at (0.8,0.3) {\large Scope};

    \draw[hc]
      (0.5,0.95) -- ($(BD.east)+(0.025,0)$)
      node[midway,left,xshift=-0.3cm,fill=white,fill opacity=0.85,text opacity=1]
      {\Large \(\bhc\)};

    % H_DL : direct-link interference CE -> Reader
    \draw[hdl]
      (0.5,0.95) -- ($(reader.north)+(0,0.03)$)
      node[midway,right,xshift=0.3cm,fill=white,fill opacity=0.85,text opacity=1]
      {\Large \(\bHdl\)};
    
    % h_R : backscatter BD -> Reader channel
    \draw[hr]
      ($(BD.east)+(0.025,0)$) -- ($(reader.north)+(0,0.03)$)
      node[midway,above,yshift=0.3cm,fill=white,fill opacity=0.85,text opacity=1]
      {\Large \(\bhr\)};

    % not a bug but to plot as last
    \node[anchor=center, box, fill opacity=0.7, yshift=-0.5cm] (CE)
      at (0.5,0.9) {\large 42 \acrlongpl{ce}};

  \end{scope}
\end{tikzpicture}

%% file: sections/system_model.tex
\section{System Model}

We consider the \gls{bibc} setup in \cref{fig:setup}. A set of $M$ phase-coherent single-antenna \glspl{ce} illuminates a single-antenna \gls{bd}, while $N$ spatially separated single-antenna readers observe the received signal.

The \glspl{ce} transmit a continuous-wave carrier $s(t)$ with beamforming vector $\bx\in\complexset{M}{1}$, subject to a per-antenna power constraint $\abs{x_m}^2\leq 1$, where $m \in \{1,\dotsc,M\}$. The relevant baseband-equivalent channels are illustrated in \cref{fig:setup} and defined as
\begin{itemize}
    \item $\bhc \in \complexset{M}{1}$: The channel between \glspl{ce} and the \gls{bd},
    \item $\bhr \in \complexset{N}{1}$: The channel between readers and the \gls{bd},
    \item $\bHbl = \bhr \bhc^\trp \in \complexset{N}{M}$: The backscatter cascade channel,
    \item $\bHdl \in \complexset{N}{M}$: The direct link %\egl{direct link?} 
    channel between \glspl{ce} and readers.
\end{itemize}
Although in the experimental setup we consider one reader, i.e., $N=1$, we use a generalized notation since the proposed algorithm also applies to multiple-reader scenarios~\cite{kaplan2025joint}.
The channels are assumed to be frequency-flat. Since the \glspl{ce} transmit single-tone continuous-wave signals, this assumption holds in practice.

The received signal at the readers contains a strong \gls{dli} component and a weak backscattered component. With reflection coefficient $\eta\, b$ (where $b$ denotes the BD modulation symbol and $\eta$ captures the reflection efficiency), the received vector can be written as

\begin{equation}
  \by_\mathrm{R}= \bHdl \sqrt{P_{\text{max}}}\, \bx\, s + \eta b\,\bHbl \sqrt{P_{\text{max}}}\, \bx\, s + \bn,
\end{equation}
where \(s\) is the transmitted symbol by the \glspl{ce}, satisfying $\abs{s}^2=0.8$, and $P_{\text{max}}$ is the transmit power of each \gls{ce}.

In this work, the performance of the \gls{bf} strategies is evaluated in terms of the relative suppression of the \gls{dli} power at the \reader while maintaining high incident power at the \gls{bd}. 

%% file: sections/algorithm.tex
% =========================
\section{Direct-Link Interference Suppression}
% =========================
This section summarizes the designed transmit \gls{bf} strategy used to suppress \gls{dli} caused by the emitters to the \reader{} while maintaining strong illumination at the \gls{bd}. This beamformer is, in the remainder of this work, referred to as the \AZF technique.

Due to the strong \gls{dli} compared to the received backscattered power at the readers, the \glspl{adc} in the readers can become saturated, or the quantization error increases while detecting the received backscattered signal. Therefore, it is important to suppress the \gls{dli} before it is received by the readers using transmit \gls{bf}. This requirement can be expressed via the \gls{sir} at the readers, defined as the ratio between the backscattered component and the \gls{dli} component. As a rule of thumb, the achievable dynamic range of an $n$-bit \gls{adc} scales linearly with $n$ (approximately $6\,n$~dB), motivating an explicit constraint on the \gls{dli} level relative to the desired backscattered signal.

In this paper, we mainly focus on the special case $\text{SIR}^{-1} =\norm{\bHdl \bx}^2/\norm{\bHbl \bx}^2 = 0$, which arises naturally in systems with one reader and multiple \glspl{ce}. In such systems, interference can be fully suppressed. More general formulations, including partially canceled \gls{dli}, are presented in~\cite{kaplan2025joint}.

\subsection{Problem Formulation}
We formulate the \gls{dli} suppression problem as 
\begin{subequations}\label{eq:original_problem}
	\begin{align}
		\mathcal{O} \mathcal{P}_{1} \quad & \underset{\bx\in \complexset{M}{1}}{\text{maximize}}
		& & \norm{\bHbl \bx}^2 \\
		& \text{subject to}
		& & \text{SIR}^{-1} = \frac{\norm{\bHdl \bx}^2}{\norm{\bHbl \bx}^2} = 0, \label{eq:dli_constraint}\\
        & & & \abs{x_m}^2 \leq 1, \forall m \in \{1,\dotsc,M\}, \label{eq:tx_power_constraint}
	\end{align}
\end{subequations}
where the constraint in \cref{eq:tx_power_constraint} limits the transmit power in each antenna. This constraint is more practical than a total power constraint, as each antenna is typically operated at maximum power and is limited by the \gls{pa} at each \gls{rf} chain rather than a total transmit power constraint.

Problem \(\mathcal{O} \mathcal{P}_{1}\) can be rewritten as
\begin{equation}\label{eq:maximization}	
\underset{\bx\in \complexset{M}{1}}{\text{maximize}}
	\abs{\bhc^\trp \bx}^2 
		 \text{subject to }
		\bHdl \bx = 0, \eqref{eq:tx_power_constraint},
\end{equation}
which follows from $\norm{\bHbl \bx}^2=\norm{\bhr}^2\abs{\bhc^\trp \bx}^2$.
This problem is non-convex due to the maximization of a norm. Let $\bx^\star$ denote an optimal solution. Then, for any arbitrary phase $\theta$, the vector $\bx^\star e^{j\theta}$ is also an optimal solution for the problem because a common phase rotation of the \gls{bf} vector does not affect the objective value or the feasibility of the constraints.
Therefore, without loss of optimality, we can assume the objective is real and non-negative, i.e., $\re{\bhc^\trp \bx} \geq 0$ and $\im{\bhc^\trp \bx} = 0$ \cite{kaplan2025joint}.
Consequently, the optimization problem can be equivalently reformulated as 
\begin{equation}\label{eq:prob_Pa0_prime_convex}
	   \underset{\bx\in \complexset{M}{1}}{\text{maximize}}
		 \re{\bhc^\trp \bx} 
		 \text{ s.t. }
	\bHdl \bx = 0,
    \im{\bhc^\trp \bx} = 0, \eqref{eq:tx_power_constraint}.
\end{equation}
The resulting formulation is convex and can be efficiently solved using convex optimization tools.
The optimal \gls{bf} vector obtained from this formulation is referred to as $\bx^\star$.

\subsection{Phase-Only Implementation}
To enforce constant transmit power across antennas, we consider a phase-only version of $\bx^\star$ obtained by keeping only the phase of each element of $\bx^\star$. The phase-only beamforming vector is defined as
\begin{equation}\label{eq:azf_phase}
x_{{\AZFmath},m}
= {x^\star_m}/{\abs{x^\star_m}},
\end{equation}
where $x_{{\AZFmath},m}$ is the $m$-th element of $\bx_{\AZFmath}$, which is the final \gls{bf} vector used for the experimental results in \cref{sec:results}.
In addition, we observe that the optimal solution $\bx^\star$ obtained from the convex formulation already exhibits nearly constant modulus across its elements, implying that the phase-only projection in \cref{eq:azf_phase} incurs only a negligible performance loss.\footnote{For further discussion on phase-only BF and its performance, see \cite{mohammed2013per}.}

%% file: figures/MRT-BD-1/heatmap.tex
% PGFPlots includegraphics example for figures/MRT-BD-1/heatmap_bitmap.png
\begin{tikzpicture}
  \begin{axis}[
    axis equal image,
    xmin=2.65857, xmax=4.00205,
    ymin=1.12411, ymax=2.37628,
    axis lines=box,
    xlabel={x [m]},
    ylabel={y [m]},
    % colorbar,
    % colormap name=inferno,
    point meta min=0,
    point meta max=172,
    enlargelimits=false,
  ]
    \addplot graphics [includegraphics cmd=\pgfimage, xmin=2.65857, xmax=4.00205, ymin=1.12411, ymax=2.37628] {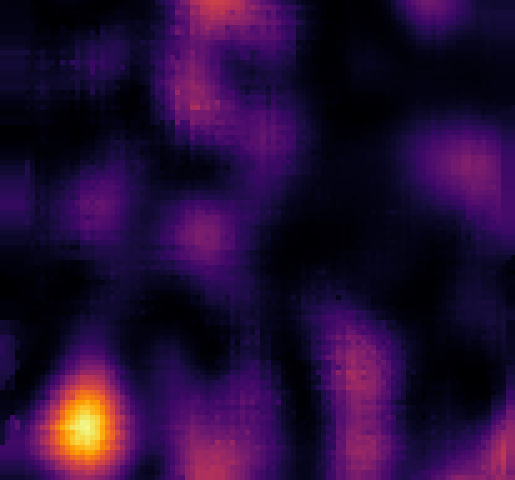};
    % Target circle (radius = wavelength/8)

    \draw[cyan,dashed,thick] (2.86,1.22642) circle [radius=0.0407609];
    \draw[cyan,dashed,thick] (2.71922,1.14) rectangle (3.0,1.38);
    \draw[cyan,dashed,thick, fill] (2.86,1.22642) circle [radius=0.002]; 
  \end{axis}
\end{tikzpicture}

%% file: figures/AZF-1/heatmap.tex
% PGFPlots includegraphics example for figures/AZF-1/heatmap_bitmap.png
\begin{tikzpicture}
  \begin{axis}[
    axis equal image,
    xmin=2.71922, xmax=3.97139,
    ymin=1.12273, ymax=2.37491,
    axis lines=box,
    xlabel={x [m]},
    colorbar,
    colorbar style={
        title={\si{\micro\watt}},
        title style={at={(1.2,0.95)}, anchor=west},
    },
    colormap name=inferno,
    point meta min=0,
    point meta max=172,
    enlargelimits=false,
  ]
    \addplot graphics [includegraphics cmd=\pgfimage, xmin=2.71922, xmax=3.97139, ymin=1.12273, ymax=2.37491] {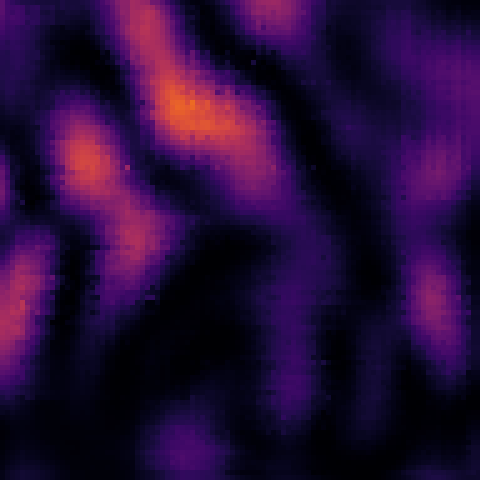};
    % Target circle (radius = wavelength/8)
    \draw[cyan,dashed,thick] (2.86,1.22642) circle [radius=0.0407609];
    \draw[cyan,dashed,thick] (2.71922,1.14) rectangle (3.0,1.38);
    \draw[cyan,dashed,thick, fill] (2.86,1.22642) circle [radius=0.002]; 
  \end{axis}
\end{tikzpicture}

%% file: figures/AZF-1/heatmap_zoom_vs_MRT-BD-1_dB.tex
% PGFPlots includegraphics example for figures/AZF-1/heatmap_zoom_vs_MRT-BD-1_dB_bitmap.png
\begin{tikzpicture}
  \begin{axis}[
    axis equal image,
    xmin=2.68466, xmax=3.02379,
    ymin=1.12411, ymax=1.39802,
    axis lines=box,
    xlabel={x [m]},
    ylabel={y [m]},
    colorbar,
    colormap name=inferno,
    point meta min=-17.8493,
    point meta max=13.4059,
    colorbar style={
        title={dB},
        title style={at={(1.2,0.95)}, anchor=west},
      ytick={-15,-10,-5,0,5,10}
    },
    enlargelimits=false,
  ]
    \addplot graphics [includegraphics cmd=\pgfimage, xmin=2.68466, xmax=3.02379, ymin=1.12411, ymax=1.39802] {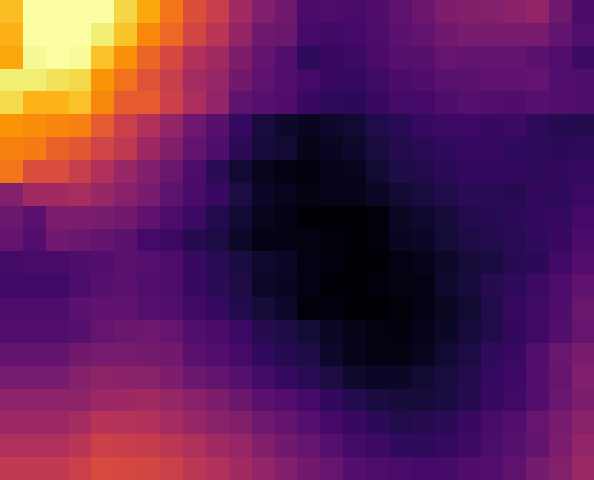};
    % Target circle (radius = wavelength/8)
    \draw[cyan,dashed,thick] (2.86,1.22642) circle [radius=0.0407609];
    \draw[cyan,dashed,thick] (2.71922,1.14) rectangle (3.0,1.38);
    \draw[cyan,dashed,thick, fill] (2.86,1.22642) circle [radius=0.002]; 

    \pgfmathsetmacro{\cx}{2.86}
    \pgfmathsetmacro{\cy}{1.22642}
    \pgfmathsetmacro{\r}{0.0407609}
    \pgfmathsetmacro{\h}{0.02} % vertical offset above the circle
    
    \draw[cyan,thick]
      (\cx-\r,\cy+\r+\h) -- (\cx+\r,\cy+\r+\h) node[midway, above] {\Large \(\lambda/8\)};;

  \end{axis}
\end{tikzpicture}

%% file: sections/exp-setup.tex
% =========================
\section{Experimental Setup and Measurement Methodology}
% =========================
The experiments were conducted using the Techtile distributed \gls{mimo} testbed~\cite{Call2206:Techtile}. The setup is shown in~\cref{fig:setup} and the experimental parameters are summarized in \cref{tab:exp_params}. 

Techtile is a modular room-scale infrastructure composed of detachable tiles on the walls, floor, and ceiling. In total, the room integrates \num{140} tiles, with \num{42} ceiling tiles that can act as a distributed transmit array. The room measures approximately \SI{4}{\meter}$\times$\SI{8}{\meter} with a height of \SI{2.4}{\meter}. Each tile integrates a processing unit and a USRP B210 \gls{sdr}, and is connected to a central controller via Ethernet for control and data aggregation~\cite{Call2206:Techtile}. The infrastructure is time, frequency and phase coherent.

\begin{table}[tb]
\centering
\caption{Experimental parameters}
\label{tab:exp_params}
\begin{tabular}{lll}
\toprule
Parameter & Variable & Value \\
\midrule
Per \Gls{ce} transmit power & \(P_\text{max}\) &\SI{11}{\dBm} \\
Number of \glspl{ce} & $M$ & \num{42} \\
\glspl{ce} total transmit power & & \SI{27.23}{\dBm} \\
\Gls{ce} antenna type && Patch (custom)\\
\Reader antenna type & & Patch (custom)\\
\BD antenna type & & Dipole (TI.92.2113)\\
Number of \reader & \(N\) & 1\\
Carrier frequency & $f_\mathrm{c}$ & \SI{920}{\mega\hertz} \\
Grid scan area & & \SI{1.25}{\meter} $\times$ \SI{1.25}{\meter} \\
\bottomrule
\end{tabular}
\end{table}

\subsection{Carrier Emitters}\label{sec:ce}
The transmit infrastructure consists of \num{42} ceiling-mounted active antenna elements arranged in a fixed grid. Each element integrates a custom-designed patch antenna with a characterized radiation pattern and an \gls{sdr}, i.e., a USRP B210. Phase coherence across the array is ensured through the calibration procedure as detailed below. \Gls{bf} weights are computed and applied using a \gls{uhd}-based Python control framework. 

Phase coherence across all \num{42} transmit channels is achieved using a multi-step calibration pipeline~\faicon{github}\footnote{\url{https://github.com/techtile-by-dramco/geometry-based-wireless-power-transfer}}. A common frequency/time reference (10\,MHz and 1\,PPS) provides a shared clock baseline, while the calibration steps below compensate the remaining per-chain phase offsets:

\textit{1. Reference \gls{rf} tone. } Each \gls{sdr} is connected to a reference cable via its RX chain. The reference \gls{rf} tone establishes an absolute phase anchor.

\textit{2. Compensation of static phase offsets. } The phase offsets between all reference \gls{rf} cables were measured manually with a scope relative to one arbitrary reference cable.

\textit{3. Loopback calibration. } TX--RX asymmetries in the \gls{lo} paths are corrected by a loopback measurement.

\subsection{Reader and Backscatter Device}

\Cref{fig:setup} shows the location of the \BD and the mobile reader,\footnote{In this paper, we consider one \reader, i.e., $N=1$.} \Cref{fig:circulator} elaborates on how the same antenna is used for pilot transmission using the \gls{sdr}, while measuring the power using an oscilloscope. Both the \reader and \gls{bd} are equipped with a circulator to support two operating modes: (i) uplink pilot transmission for channel estimation, and (ii) downlink power measurement during transmit \gls{bf}. In pilot mode, the antenna is connected to a USRP B210 that transmits a single-tone \gls{cw} pilot. In measurement mode, the circulator routes the received \gls{rf} signal to a calibrated oscilloscope, which reports the received power. Hence, the \BD is not operated as a passive backscatter device but is actively emulated using measurement instrumentation. In the remainder of this work, this emulated backscatter device is referred to as the \BD.

\begin{figure}[hbpt]
    \centering
    \resizebox{0.8\linewidth}{!}{%
        \input{figures/circulator.tikz}%
    }
    \caption{Dual pilot and received power measurement setup. The \gls{rf} circulator is used to transmit the pilot via the uplink and route the downlink power to the oscilloscope for calibrated power measurements. This allows us to send a pilot signal through the \gls{sdr}, while receiving the power at the oscilloscope.}\label{fig:circulator}
\end{figure}

To map the spatial distribution of interfering power at and around the \reader location, the receiver is mounted on a motorized XY-plotter that scans a two-dimensional grid within the measurement area of \SI{1.25}{\meter}~$\times$~\SI{1.25}{\meter}.

Ground-truth positioning is obtained using a Qualisys motion capture system. Retro-reflective markers mounted on the \reader enable three-dimensional tracking with millimeter-level accuracy. All positions reported in this work are expressed in the coordinate frame of the motion capture system.

\subsection{Experiment Procedure}

The experiment procedure is as follows:

\textit{1. Array synchronization and calibration. }
The \glspl{ce} are phase- and time-synchronized. To exploit channel reciprocity, the analog front-end is calibrated because the TX and RX RF chains are not reciprocal and use separate \glspl{lo}. A loopback procedure is executed to estimate and compensate the phase ambiguity between the RX and TX chains, as detailed in \cref{sec:ce}.

\textit{2. Channel sounding. } 
The \BD and \reader sequentially transmit a single-tone \gls{cw} pilot. During this step, the \glspl{ce} receive the pilots and estimate the required \gls{csi}, i.e., $\bhc \in \complexset{M}{1}$ and $\bhdl \in \complexset{1}{M}$. Note that $\bhr$ is not required because we only measure the received power at the \gls{bd} and the \gls{dli} power in the \reader. In addition, the proposed \gls{bf} algorithm does not rely on the knowledge of $\bhr$ as shown in Eq. \eqref{eq:maximization}.

\textit{3. Beamformer computation and distribution. }
The estimated \gls{csi} is sent to a central server that computes \gls{bf} weights for the selected strategy (\gls{pomrt} or \AZF). The weights are then distributed back to the \glspl{ce}.

\textit{4. Downlink transmission and measurements. }
The \glspl{ce} apply the \gls{bf} weights and transmit the downlink carrier. We record the received RF power at both the \BD and the \reader (via the calibrated oscilloscope). For the heatmaps, the \reader is moved across the XY grid during Step~4, without altering the beamforming weights. Therefore the targeted \reader location is kept, while the physical \reader is repurposed to measure the spatial received power around the target \reader location.

\subsection{Performance Metrics}

We evaluate the performance of the \gls{bf} strategies using three power-based metrics: (i) \textit{interference suppression}, defined as the reduction in received power at the \reader location when using \AZF compared to \gls{pomrt}; (ii) \textit{BD illumination}, quantified by the received power at the \BD, \(P_{\mathrm{BD}}\); and (iii) \textit{power ratio}, defined as \(\Delta=P_{\mathrm{BD}}/P_{\mathrm{R}}\), where \(P_{\mathrm{R}}\) denotes the received power at the \reader. Since the received backscattered power at the \reader scales with the incident carrier power at the \BD, \(P_{\mathrm{BD}}\) serves as a proxy for the desired backscatter strength. For the single-\reader case and fixed tag reflection efficiency and BD--\reader channel, the backscatter \gls{sir} scales as \(\mathrm{SIR}\propto \abs{\bhc^\trp \bx}^2/\norm{\bHdl \bx}^2\), 
where  $\abs{\bhc^\trp \bx}^2$ and \(\norm{\bHdl \bx}^2\) are proportional to \(P_{\mathrm{BD}}\) and \(P_{\mathrm{R}}\), respectively.
Hence, \(\Delta\) provides a practical proxy for the expected \gls{sir} improvement and the \gls{adc} headroom.

%% file: figures/circulator.tikz
% Antenna symbol (scaled to match component proportions)
\def\antenna{%
  -- +(0mm,8.0mm)
  -- +(6.4mm,18.4mm)
  -- +(-6.4mm,18.4mm)
  -- +(0mm,8.0mm)
}

\begin{tikzpicture}[
  >=Latex,
  box/.style={
    draw,
    rectangle,
    fill=gray!15,
    minimum width=2.0cm,
    minimum height=1.0cm,
    align=center,
    outer sep=0,
  },
  circ/.style={
    draw,
    circle,
    minimum size=2.0cm,
    line width=0.8pt,
    outer sep=0,
  },
  wire/.style={line width=0.8pt}
]

% --- Nodes ---
\node[box, minimum width=3.0cm] (b210) {B210\\Pilot Source};

\node[circ, right=1.8cm of b210] (circulator) {};

% Antenna feed point
\coordinate (ant) at ($(circulator.east)+(1.6cm,0)$);

\node[box, below=1.6cm of circulator, minimum width=3.0cm] (scope) {Oscilloscope};

\draw[wire, ->, green!60!black]
  ($(b210.east)+(2mm,2mm)$)
  -- node[pos=0.2, yshift=1mm, anchor=south]{Pilot}
   ($(ant)+(-7mm,2mm)$)
  -- ++(0,1cm);

  \draw[wire, ->, green!60!black]
  ($(ant)+(-7mm,-2mm)$)
    -- ($(circulator.center)+(2mm,-2mm)$)
    -- node[pos=0.5, xshift=1mm, anchor=west]{RX Power}
    ($(scope.north)+(2mm,2mm)$);

% --- Circulator rotation arrow ---
\draw[wire, ->] ($(circulator.center)+(-0.42,0.42)$)
  arc[start angle=135, end angle=-45, radius=0.55cm];

% --- Connections ---
\draw[wire] (b210.east) -- (circulator.west);
\draw[wire] (circulator.east) -- (ant);
\draw[wire] (circulator.south) -- (scope.north);

% --- Antenna drawing ---
\draw[wire] (ant) \antenna;
\node[right=2mm of ant] {Antenna};

% --- Port labels ---
\node[above=1.5mm of $(b210.east)!0.9!(circulator.west)$] {1};
\node[above=1.5mm of $(circulator.east)!0.1!(ant)$] {2};
\node[left=1.5mm of $(circulator.south)!0.1!(scope.north)$] {3};

\end{tikzpicture}

%% file: figures/power-meas-high.tikz
\begin{tikzpicture}

\pgfplotstableread[col sep=comma, trim cells=true]
{figures/power_measurements.csv}\powertable

% Left y-axis: Reader + BD powers (dBm)
\begin{axis}[
  width=\linewidth,
  height=4cm,
  xlabel={Number of active \glspl{ce}},
  ylabel={Power (dBm)},
  grid=both,
  xmin=9, xmax=43,
  ymin=-53.5,ymax=2,
  xtick={10,15,20,25,30,35,40,42},
  axis y line*=left,
  axis x line*=bottom,
]

% Reader
% Reader (PR)
\addplot[color=rcolor, solid, mark=triangle*, mark color=rcolor]
  table[
    x=Antennas,
    y expr=\thisrow{PR_AZF}+\PRoffset
  ]{\powertable};
%\addlegendentry{Reader AZF}

\addplot[color=rcolor, dashed, mark=triangle*, mark color=rcolor]
  table[
    x=Antennas,
    y expr=\thisrow{PR_MRT}+\PRoffset
  ]{\powertable};
%\addlegendentry{Reader MRT}

% BD
\addplot[color=bdcolor, solid, mark=*, mark color=bdcolor]
  table[
    x=Antennas,
    y expr=\thisrow{PBD_AZF}+\BDoffset
  ]{\powertable};
%\addlegendentry{BD AZF}

\addplot[color=bdcolor, dashed, mark=*, mark color=bdcolor]
  table[
    x=Antennas,
    y expr=\thisrow{PBD_MRT}+\BDoffset
  ]{\powertable};

%\addlegendentry{BD MRT}

\end{axis}

%--- Minimal 2-row legend on top of the axis ---
\node[
  anchor=south,
  fill=white,
  fill opacity=0.85,
  text opacity=1,
  inner sep=3pt,
  draw=none
] at (rel axis cs:0.5,1.04) {%
\scriptsize
\begin{tabular}{@{}lll@{}}
\tikz[baseline=-0.6ex]{%
  %\draw[color=rcolor,line width=5pt] (0,0)--(1.2em,0);
  \filldraw[fill=rcolor,draw=white,line width=0.4pt] (0.6em,3pt) -- ++(-3pt,-6pt) -- ++(6pt,0) -- cycle;
}
$P_{\mathrm{R}}$ (Reader) & \tikz{\draw[black,solid,line width=0.9pt] (0,0)--(1.2em,0);} 
\AZF \\

\tikz[baseline=-0.6ex]{%
  %\draw[color=bdcolor,line width=5pt] (0,0)--(1.2em,0);
  \filldraw[fill=bdcolor,draw=white,line width=0.4pt] (0.6em,0) circle (3pt);
}
$P_{\mathrm{BD}}$ (BD) &
\tikz{\draw[black,dashed,line width=0.9pt] (0,0)--(1.2em,0);} 
MRT\\
\end{tabular}
};

\end{tikzpicture}

%% file: figures/power-meas-low.tikz
\begin{tikzpicture}

\pgfplotstableread[col sep=comma, trim cells=true]
{figures/power_measurements.csv}\powertable

% Left y-axis: Reader + BD powers (dBm)
\begin{axis}[
  width=\linewidth,
  height=4cm,
  xlabel={Number of active \glspl{ce}},
  grid=both,
  xmin=9, xmax=43,
  ymin=-53.5,ymax=2,
  xtick={10,15,20,25,30,35,40,42},
  axis y line*=left,
  axis x line*=bottom,
]

\def\PRoffset{10.67}
\def\BDoffset{8.45}

% Reader AZF WORST
\addplot[color=rcolor, solid, mark=triangle*, mark color=rcolor]
  table[
    x=Antennas,
    y expr=\thisrow{PR_AZF_WORST}+\PRoffset
  ]{\powertable};

% Reader MRT WORST
\addplot[color=rcolor, dashed, mark=triangle*, mark color=rcolor]
  table[
    x=Antennas,
    y expr=\thisrow{PR_MRT_WORST}+\PRoffset
  ]{\powertable};

% BD AZF WORST
\addplot[color=bdcolor, solid, mark=*, mark color=bdcolor]
  table[
    x=Antennas,
    y expr=\thisrow{PBD_AZF_WORST}+\BDoffset
  ]{\powertable};

% BD MRT WORST
\addplot[color=bdcolor, dashed, mark=*, mark color=bdcolor]
  table[
    x=Antennas,
    y expr=\thisrow{PBD_MRT_WORST}+\BDoffset
  ]{\powertable};

% \addlegendentry{BD MRTWORST}

\end{axis}

% --- Minimal 2-row legend on top of the axis ---
\node[
  anchor=south,
  fill=white,
  fill opacity=0.85,
  text opacity=1,
  inner sep=3pt,
  draw=none
] at (rel axis cs:0.5,1.04) {%
\scriptsize
\begin{tabular}{@{}lll@{}}
\tikz[baseline=-0.6ex]{%
  %\draw[color=rcolor,line width=5pt] (0,0)--(1.2em,0);
  \filldraw[fill=rcolor,draw=white,line width=0.4pt] (0.6em,3pt) -- ++(-3pt,-6pt) -- ++(6pt,0) -- cycle;
}
$P_{\mathrm{R}}$ (Reader) & \tikz{\draw[black,solid,line width=0.9pt] (0,0)--(1.2em,0);} 
\AZF \\

\tikz[baseline=-0.6ex]{%
  %\draw[color=bdcolor,line width=5pt] (0,0)--(1.2em,0);
  \filldraw[fill=bdcolor,draw=white,line width=0.4pt] (0.6em,0) circle (3pt);
}
$P_{\mathrm{BD}}$ (BD) &
\tikz{\draw[black,dashed,line width=0.9pt] (0,0)--(1.2em,0);} 
MRT\\
\end{tabular}
};

\end{tikzpicture}

%% file: figures/power-meas-delta.tikz
\begin{tikzpicture}
\pgfplotstableread[col sep=comma, trim cells=true]
{figures/power_measurements.csv}\powertable
% Right y-axis: Delta (dB)
\begin{axis}[
  width=\linewidth,
  height=4cm,
  xmin=9, xmax=43,
 xtick={10,15,20,25,30,35,40,42},
  axis y line*=left,
  axis x line*=bottom,
  ylabel={$\Delta$ (dB)},
  xlabel={Number of active antennas},
  legend style={at={(0.98,0.98)}, anchor=north east},
]

\addplot[black, solid, mark=pentagon*]
  table[
    x=Antennas,
    y expr=\thisrow{Delta_AZF}+\BDoffset-\PRoffset
  ]{\powertable};
% \addlegendentry{$\Delta$ AZF}

\addplot[black, dashed, mark=diamond*]
  table[
    x=Antennas,
    y expr=\thisrow{Delta_MRT}+\BDoffset-\PRoffset
  ]{\powertable};
% \addlegendentry{$\Delta$ MRT}

\addplot[black!30, solid, mark=pentagon*]
  table[
    x=Antennas,
    y expr=\thisrow{Delta_AZF_WORST}+\BDoffset-\PRoffset
  ]{\powertable};
% \addlegendentry{$\Delta$ AZF}

\addplot[black!30, dashed, mark=diamond*]
  table[
    x=Antennas,
    y expr=\thisrow{Delta_MRT_WORST}+\BDoffset-\PRoffset
  ]{\powertable};
% \addlegendentry{$\Delta$ MRT}

\end{axis}

% --- Minimal 2-row legend on top of the axis ---
\node[
  anchor=south,
  fill=white,
  fill opacity=0.85,
  text opacity=1,
  inner sep=3pt,
  draw=none
] at (rel axis cs:0.5,1.04) {%
\scriptsize
\begin{tabular}{@{}ll@{}}
\tikz[baseline=-0.6ex]{%
  \draw[black,solid,line width=0.9pt] (0,0)--(1.2em,0);
  %\draw[black] plot[mark=pentagon*,mark size=2.2pt] coordinates {(0.6em,0)};
} 
\AZF & \tikz{\draw[black,opacity=1.0,line width=0.9pt] (0,0)--(1.2em,0);} 
high$\rightarrow$low $\lvert h_{\mathrm{BD},m}\rvert^2$\\

\tikz[baseline=-0.6ex]{%
  \draw[black,dashed,line width=0.9pt] (0,0)--(1.2em,0);
  %\draw[black] plot[mark=diamond*,mark size=2.2pt] coordinates {(0.6em,0)};
} 
MRT &
\tikz{\draw[black,opacity=0.35,line width=0.9pt] (0,0)--(1.2em,0);} 
low$\rightarrow$high $\lvert h_{\mathrm{BD},m}\rvert^2$\\
\end{tabular}
};

\end{tikzpicture}

%% file: sections/exp-results.tex
% =========================
\section{Experiment Results}
% =========================
\label{sec:results}
We compare baseline \gls{pomrt} beamforming (maximizing illumination at the \BD) against the proposed \AZF beamforming, which enforces a spatial null at the \reader location while maintaining \BD illumination. Both strategies use the same per-antenna transmit power budget. 
\AZF refers to the phase-only implementation in \cref{eq:azf_phase}.
For \gls{pomrt}, the $m$-th element of the \gls{bf} vector is designed as 
\begin{equation}\label{eq:mrt}
    x_{\text{po-mrt}, m}
= {\matr{h}_{\text{C},m}^*}/{\abs{\matr{h}_{\text{C},m}}}.
\end{equation}

\subsection{Direct-Link Interference Spatial Heatmaps}

\Cref{fig:spatial-heatmaps-beamforming} visualizes the spatial distribution of the received power over the entire measurement area when \gls{bf} is performed toward the \BD using \num{42} \glspl{ce} using the \gls{pomrt} (\cref{fig:heatmap-mrt-bd}) and \AZF (\cref{fig:heatmap-azf}) beamformer. The first heatmap corresponds to \gls{pomrt} optimizing power transfer to the \BD without any notion of a reader location. As a result, all spatial positions are potential reader locations, and significant \gls{dli} is observed at multiple points across the area (\cref{fig:heatmap-mrt-bd}).

In contrast, \AZF explicitly accounts for the estimated reader channel by imposing a spatial null at the reader location while maintaining strong illumination at the \BD (\cref{fig:heatmap-azf}). At the target reader location (indicated by the dashed circle)\footnote{This location is intentionally chosen because it corresponds to a local maximum of the \gls{dli} under \gls{pomrt}, enabling a clear evaluation of the nulling capability of the \AZF strategy.}, the measured \gls{dli} is reduced from \SI{-8.06}{\dBm} under \gls{pomrt} to \SI{-25.30}{\dBm} with \AZF, corresponding to approximately \SI{17}{\decibel} of local suppression.

\Cref{fig:heatmap-azf-vs-mrt} shows the ratio, in \si{\decibel}, between the received power at each spatial location under \AZF and \gls{pomrt}. The zoomed-in differential heatmap confirms that the suppression is highly localized around the intended reader position, illustrating both the spatial selectivity and the placement sensitivity of null steering with a finite-size array. A small spatial offset between the intended and realized null location (approximately $\lambda/30$) is observed, which is attributed to residual calibration errors and/or environmental dynamics.

\subsection{Performance with Different Number of Carrier Emitters}
To quantify how performance scales with the number of active transmit elements and \glspl{ce} selection, we form sub-arrays by selecting $K\in\{10,20,30,40,42\}$ \glspl{ce}. We order the \glspl{ce} based on the estimated BD-link channel power $\lvert h_{\mathrm{BD},m}\rvert^2$ and consider two selection rules: high$\rightarrow$low (strongest-first) and low$\rightarrow$high (weakest-first). For each $K$, the measured received power at the \reader, $P_{\mathrm{R}}$, and at the \BD, $P_{\mathrm{BD}}$, are shown in Figures \ref{fig:power-meas-high} and \ref{fig:power-meas-low}. In addition, we report the power ratio metric $\Delta=P_{\mathrm{BD}}/P_{\mathrm{R}}$ in \cref{fig:power-meas-delta}. The reader is located at the same position as in the heatmap experiments, depicted by the circle in \cref{fig:spatial-heatmaps-beamforming}.

\Cref{fig:power-meas} shows that \gls{pomrt} provides limited separation between \BD illumination and \reader interference: increasing $K$ boosts $P_{\mathrm{BD}}$ but also raises $P_{\mathrm{R}}$, yielding $\Delta\approx\SIrange{6.5}{8.5}{\decibel}$ for strongest-first selection. In contrast, \AZF keeps $P_{\mathrm{R}}$ substantially lower while maintaining high $P_{\mathrm{BD}}$, giving $\Delta\approx\SIrange{25}{39}{\decibel}$ and improving ratio by up to \SI{31}{\decibel} over \gls{pomrt} (e.g., \SI{39.09}{\decibel} vs.\ \SI{8.49}{\decibel} at $K=20$). Even under weakest-first selection, \AZF maintains a clear advantage ($\Delta\approx\SIrange{11.6}{25.1}{\decibel}$), while \gls{pomrt} remains within about \SIrange{-4.8}{5.4}{\decibel}. At $K=42$, switching from \gls{pomrt} to \AZF reduces $P_{\mathrm{R}}$ by about \SI{18.6}{\decibel} while $P_{\mathrm{BD}}$ drops by only \SI{0.64}{\decibel} (\SI{-13.14}{\dBm} to \SI{-13.78}{\dBm}).